\documentclass[useAMS]{mn2e}
\usepackage{latexsym,graphicx}
\newcommand{\solm}{M$_{\odot}$}
\usepackage{stfloats}

\title[Near-infrared polarimetry as a tool ...]
{Near-infrared polarimetry as a tool for testing properties of accreting super-massive black holes}
\author[Zamaninasab et al.]{M. Zamaninasab$^{1,2}$\thanks{E-mail:
zamani@mpifr-bonn.mpg.de}, A.~Eckart$^{2,1}$, M.~Dov\v{c}iak$^{3}$, V.~Karas$^{3}$, R.~Sch\"odel$^{4}$, G.~Witzel$^{2}$,\
\newauthor
 N.~Sabha$^{2}$, M.~Garc\'{\i}a-Mar\'{\i}n$^{2}$, D.~Kunneriath$^{2,1}$, K.~Mu\v{z}i\'{c}$^{5}$, C.~Straubmeier$^{2}$,\
\newauthor
M. Valencia-S$^{2,1}$ and J.A.~Zensus$^{1,2}$\\
$^{1}$Max-Planck-Institut f\"ur Radioastronomie, Auf dem H\"ugel 69, 53121 Bonn, Germany\\
$^{2}$I.Physikalisches Institut, Universit\"at zu K\"oln, Z\"ulpicher Str.77, 50937 K\"oln, Germany\\
$^{3}$Astronomical Institute, Academy of Sciences, Bo\v{c}n\'{i} II 1401, CZ-14131 Prague, Czech Republic\\
$^{4}$Instituto de Astrof\'{\i}sica de Andaluc\'{\i}a - CSIC, Glorieta de la Astronomía S/N, 18008 Granada, Spain\\
$^{5}$University of Toronto, Dept. of Astronomy and Astrophysics 50 St. George St. Toronto, ON M5S 3H4, Canada}

\begin{document}

\date{Accepted 2010 December 1. }

\pagerange{\pageref{firstpage}--\pageref{lastpage}} \pubyear{2010}

\maketitle

\label{firstpage}

\begin{abstract}
Several massive black holes exhibit flux variability on time scales that correspond to source
sizes of the order of few Schwarzschild radii. We survey the potential of near-infrared and X-ray 
polarimetry to constrain physical properties of such black hole systems, namely their spin
and inclination. We have focused on a model where an orbiting hot spot is embedded in an accretion disk.
A new method of searching for the time-lag between orthogonal polarization channels is developed and 
applied to an ensemble of hot spot models that samples a wide range of parameter space. 
We found that the hot spot model predicts signatures in polarized light which are in the range 
to be measured directly in the near future.
However, our estimations are predicted upon the
assumption of a Keplerian velocity distribution inside the 
flow where the dominant part of the magnetic field is toroidal.  

We also found that
if the right model of the accretion flow can be chosen for each source 
(e.g. on the basis of magnetohydrodynamics simulations)
then the black hole spin and inclination can be constrained to a small two-dimensional area in the spin-inclination space.
The results of the application of the method to the available near-infrared polarimetric data of Sgr~A* is  
presented. It is shown that even with the currently available data the 
spin and inclination of Sgr~A* can be constrained.
 Next generations of near-infrared and X-ray polarimeters 
should be able to exploit this tool.
\end{abstract}

\begin{keywords}
black hole physics: general, infrared: general, accretion,
accretion disks, Galaxy: centre, Galaxy: nucleus
\end{keywords}


\begin{figure}
\centering{\includegraphics[width=0.45\textwidth]{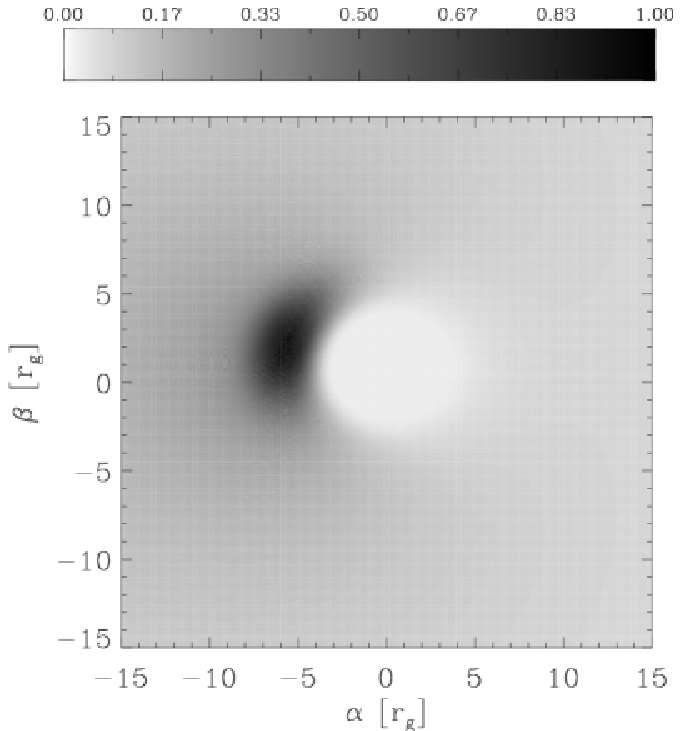}}
\centering{\includegraphics[width=0.45\textwidth]{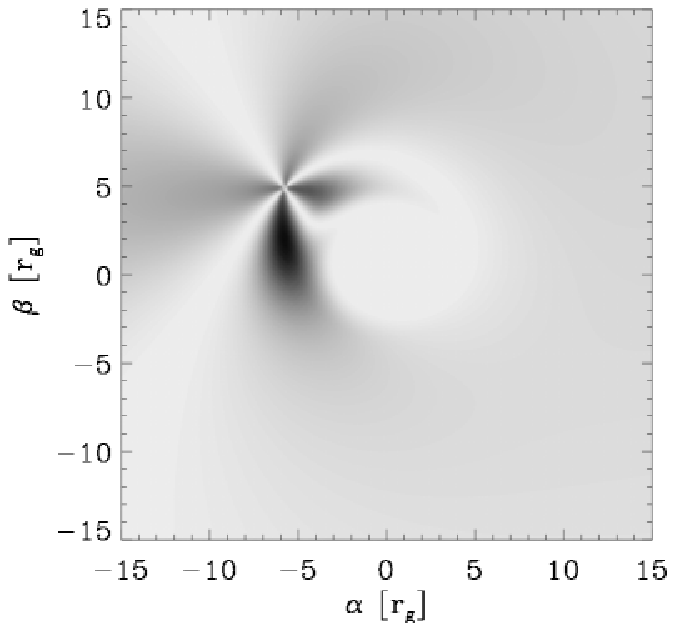}}
\centering{\includegraphics[width=0.45\textwidth]{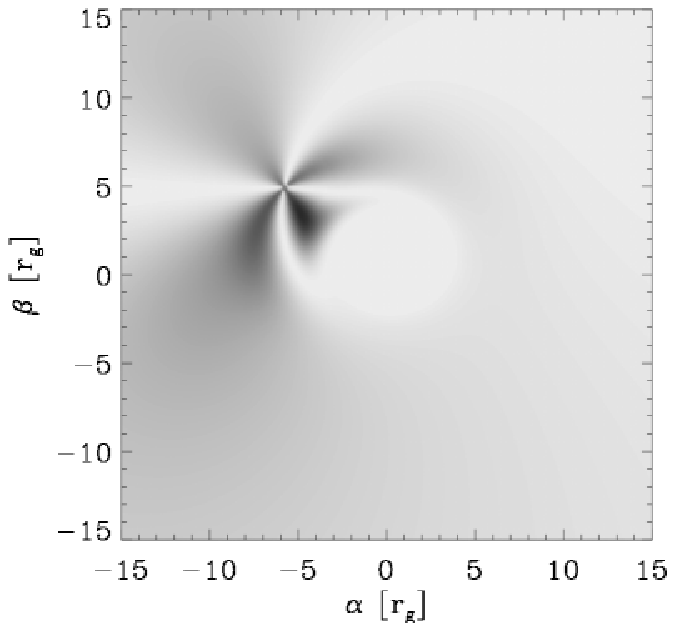}}
\caption[]{Apparent images of an orbiting spot inside the accretion flow around 
a Kerr black hole with the spin $a=0.5$ viewed by an observer inclined by $35^\circ$.
The image on top shows the ray-tracing output for the total flux, while the middle 
and bottom show the flux in the $0^\circ$ and $90^\circ$ polarization channels, respectively.
Here we have assumed that the projected spin axis is aligned toward the north
on the observer's sky ($\theta=0^\circ$)}.
\label{image}
\end{figure}

\section{Introduction}
The high-frequency emission from black-hole systems originates mostly close to the black-hole 
horizon. In this central engine, huge amount of gravitational potential energy is 
released from gas falling into the black hole. It is known that this emission can vary extremely 
rapidly (e.g. see Gaskell et al. 2007). 
This variable emission from compact objects has been interpreted by several authors
as being related to the presence of orbiting spots in the inner parts of the
 accretion flows (Karas \& Bao 1992; Karas 1999; Shahbaz 1999; Stella \& Vietri 1999; Dov\v{c}iak et al. 2004; Schnittman et al. 2006; Broderick \& Loeb 2006, Zamaninasab et al. 2010).
Here we will focus on a new method based on the high-frequency time-resolved polarimetric 
observations which can 
constrain the intrinsic parameters of the sources.

In the case of stellar-mass black hole candidates, 
the resemblance between the quasi-periodic oscillation (QPO) 
time scales and the dynamical frequencies of the black holes near their 
innermost stable circular orbit (ISCO) has inspired the \textit{hot spot} model 
(Stella \& Vietri 1998, 1999). 
An orbiting blob inside the accretion flow will produce 
observable effects in both total and polarized flux. 
In the case of super-massive black holes, the situation is even more challenging 
 because the measured fluxes are much lower and time scales are much longer.
Nonetheless, it is expected that similar physical mechanisms act in both categories of 
accreting black holes and, after appropriate rescaling, can be considered by similar techniques (Karas and Matt 2007).
In Zamaninasab et al (2010),  we have discussed in detail how 
flux magnification and changes in polarization angle (degree) follow specific 
patterns in this scenario, and we probed such a pattern in a sample of
the near-infrared (NIR) flares of Sgr~A*.

Several mechanisms have been proposed for the creation of the hot spots,
including the random variations resulted from magnetic turbulence inside the 
magnetohydrodynamics (MHD) flow (Balbus \& Hawley 1991, Armitage \& Reynolds 2003), vortices and 
flux tubes (Abramowicz et al. 1992), magnetic flares (Poutanen \& Fabian 1999, Zycki 2002)
 or star-disk interactions (Nayakshin et al. 2004, Dai et al. 2010).
It is also shown that  multi-component spot scenarios are able to reproduce
the  overall behaviour of the observed power spectral densities (PSDs) and
their (transient) QPO features (Schnittman et al. 2006, Pech\'{a}\v{c}ek et al. 2008, Zamaninasab et al. 2010).

In this paper, we will discuss a straightforward method which can be used for 
constraining the physical parameters of black holes (or other compact objects) which show 
variable polarized activity.
The new method has several advantages in comparison with previous works
(Meyer et al. 2006a,b, 2007, Hamaus et al. 2009, Zamaninasab et al. 2008a,b, 2010).
We must note that this method, in its core, assumes that variabilities
in the flux are caused by orbiting spots. Normally, it has
been reported in the literature, that such spots happen to
be located at the ISCO. Here we will consider other
possibilities for the spot's location and show how they will
affect the predictions of the model (see Dov\v{c}iak et al. 2004, Broderick \& Loeb 2006 and Reid et al. 2008
for detailed discussion about 
the spot model and effect of the position of the spot on the observed signal).
We must also stress that the results derived here are dependent 
on our assumptions about the Keplerian
velocity of the flow and the magnetic field orientation.

\begin{figure}
\centering{\includegraphics[width=0.48\textwidth]{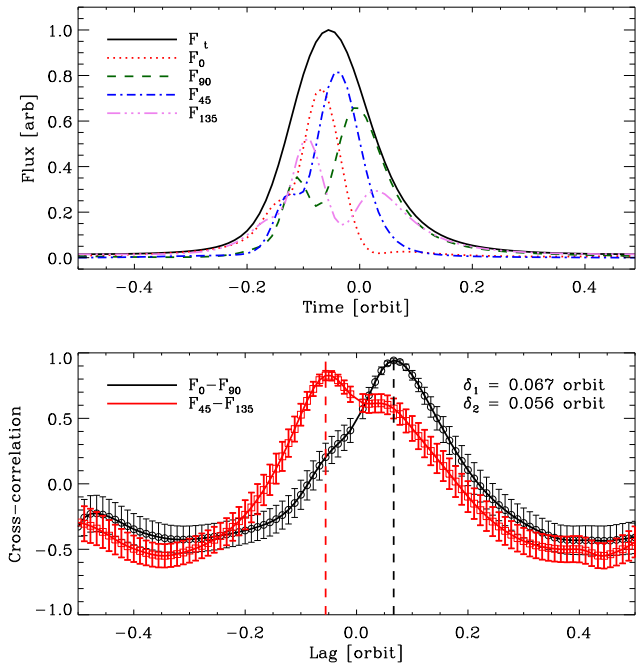}}
\caption[]{Top: Total flux magnification and the corresponding 
polarization channels  for a spot orbiting at the marginally stable orbit of 
a Schwarzschild black hole and viewed by an observer inclined by $60^\circ$.
Flux densities are all normalized to the maximum of the total flux value. 
Bottom: Cross-correlation between light curves of orthogonal channels: 
$0^\circ$ \& $90^\circ$ (black) and $45^\circ$  \& $135^\circ$ (red).
The cross-correlation shows a clear delay between orthogonal pairs 
of channels with the values of $\delta_1$ and $\delta_2$. 
Here we have assumed that the projected spin axis is aligned toward the north
on the observer's sky ($\theta=0^\circ$)}.
\label{lag}
\end{figure}

In section \ref{time_delay_sect}
we will describe how the spot model predicts a time-lag between orthogonal polarization channels
and study how these time-lags are characterized by the black hole mass, spin and inclination angle.
A description about the available NIR data from the Galactic Centre black hole and how they 
can be used to constrain its spin and inclination
is provided in section \ref{sgra_sect}. We also briefly discuss a possible way to apply this method 
 to other sources, like RE~J1034+396. In section \ref{conclusion},
 we summarize the main results and draw
our conclusions.

\section{Time Delay Between Polarization Channels: Predictions of the Hot Spot Model}
\label{time_delay_sect}

MHD simulations have 
demonstrated that in the inner parts of the accreting plasmas,
 magnetic-field dissipation moderates the field intensity far below its
 equipartition value (Kowalenko \& Melia 1999). Nevertheless,
 when the gas propagates and follows a Keplerian flow, a MHD dynamo 
can generate an increased (yet subequipartition) magnetic field that
 is dominated by its azimuthal component (Hawley, Gammie, \& Balbus 1996).
In this paper we examine fluctuations occurring in the inner parts of
the black hole's accretion flow, very close to the event horizon. 
Since our focus is mainly on radii smaller than $40 \ r_g$ (where $r_g=\frac{GM}{c^2}$ 
is the gravitational radius) we have adopted a thin disk approximation while
magnetic field is dominated by its azimuthal component.
Thin disk approximation was also chosen for reasons of practicability.
It allows in a straight forward way (low computational effort)
to repeat simulations for a wide range of free-parameters space.
Adopting a relativistic treatment of the inner Keplerian region, 
while considering certain aspects of earlier hydrodynamical simulations, 
we end up with a model where the 
gas circulates at small radii twisting the magnetic field inside a relativistic Keplerian disk. 
To estimate the possible structure of the gas distribution within several Schwarzschild 
radii of the black hole, we followed the investigation done by Qian et al. (2009).

For the radiation mechanism, we assumed that NIR 
and X-ray synchrotron photons originate from the relativistic electrons 
gyrating around magnetic-field lines inside the flow 
(see Liu et al. 2006, Eckart et al. 2004, 2006, 2008, Zamaninasab et al. 2010).
The energy distribution
of electrons is approximated by a power-law function:
\begin{equation}
N(\gamma)= \left\{\begin{array}{cc}{{N_0\gamma^{-p}}} \  \,
& \gamma \leq \gamma_c \\
    \\
    {{0 }} \ \, &
    \gamma > \gamma_c \end{array}\right.
\label{elec_dist}
\end{equation}
where $N(\gamma), \gamma$ and  $\gamma_c$ are electron energy distribution function,
Lorentz factor of the electrons and its cut-off, respectively.
Thus we can define both the ordinary
and extraordinary emission coefficients for each point of
the accretion disk (Pacholczyk 1970; see also Zamaninasab et al. 2010):
\begin{equation}
\epsilon_\nu^{\pm} \ = \frac{\sqrt{3}e^3}{8\pi m_ec^2}\ N_0 \ B\sin\theta_e \int{\gamma^{-p} [F(x)\pm G(x)]d\gamma }
\end{equation}
where 
\begin{eqnarray}
F(x)&=&\int^{\infty}_{x} {K_{5/3}(z)dz}\\
G(x)&=&xK_{2/3}(x)\\
x&=&\frac{4\pi m_e^3c^5 \nu}{3eB\sin{\theta_e}\gamma^2}
\end{eqnarray}
$K_{5/3}$ and $K_{2/3}$ are the corresponding modified Bessel functions,
$\nu$ is the frequency measured in the
co-moving frame, $B$ is the magnetic-field strength, and $\theta_e$ is the angle between the direction of the
magnetic field and the direction toward the co-moving observer:
\begin{equation}
\cos(\theta_e)={\sqrt{\frac{({B}^\alpha p_{e\alpha})^2}{(p_e^\beta p_{e\beta})({B}^\gamma
{B}_\gamma)}}} 
\end{equation}
In the above equation $B$ is the magnetic field four-vector and $p_e$ represents the direction of the emitted photon momentum 
in the co-moving frame.

\begin{figure*}
\centering{\includegraphics[width=0.33\textwidth]{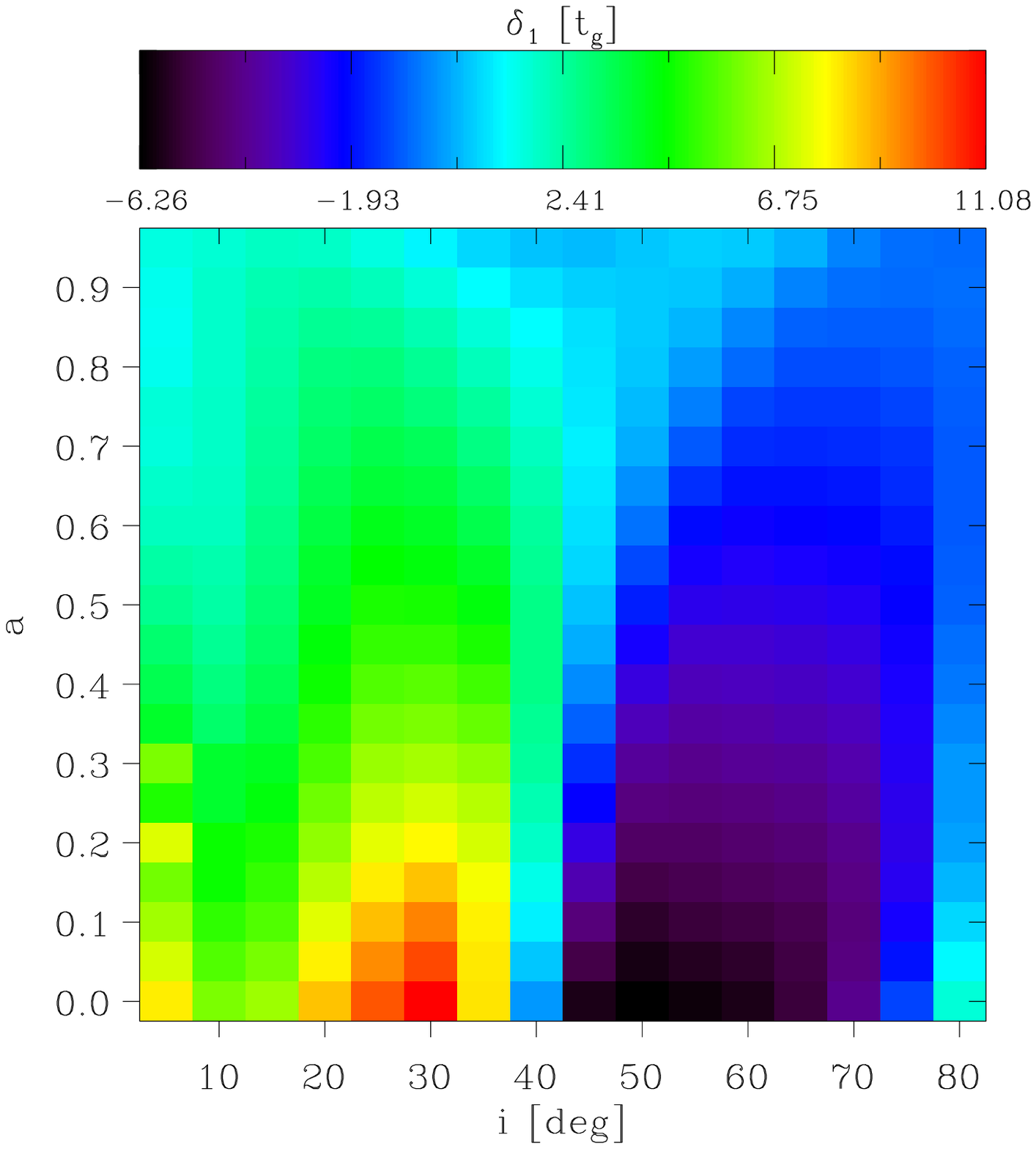}}
\centering{\includegraphics[width=0.33\textwidth]{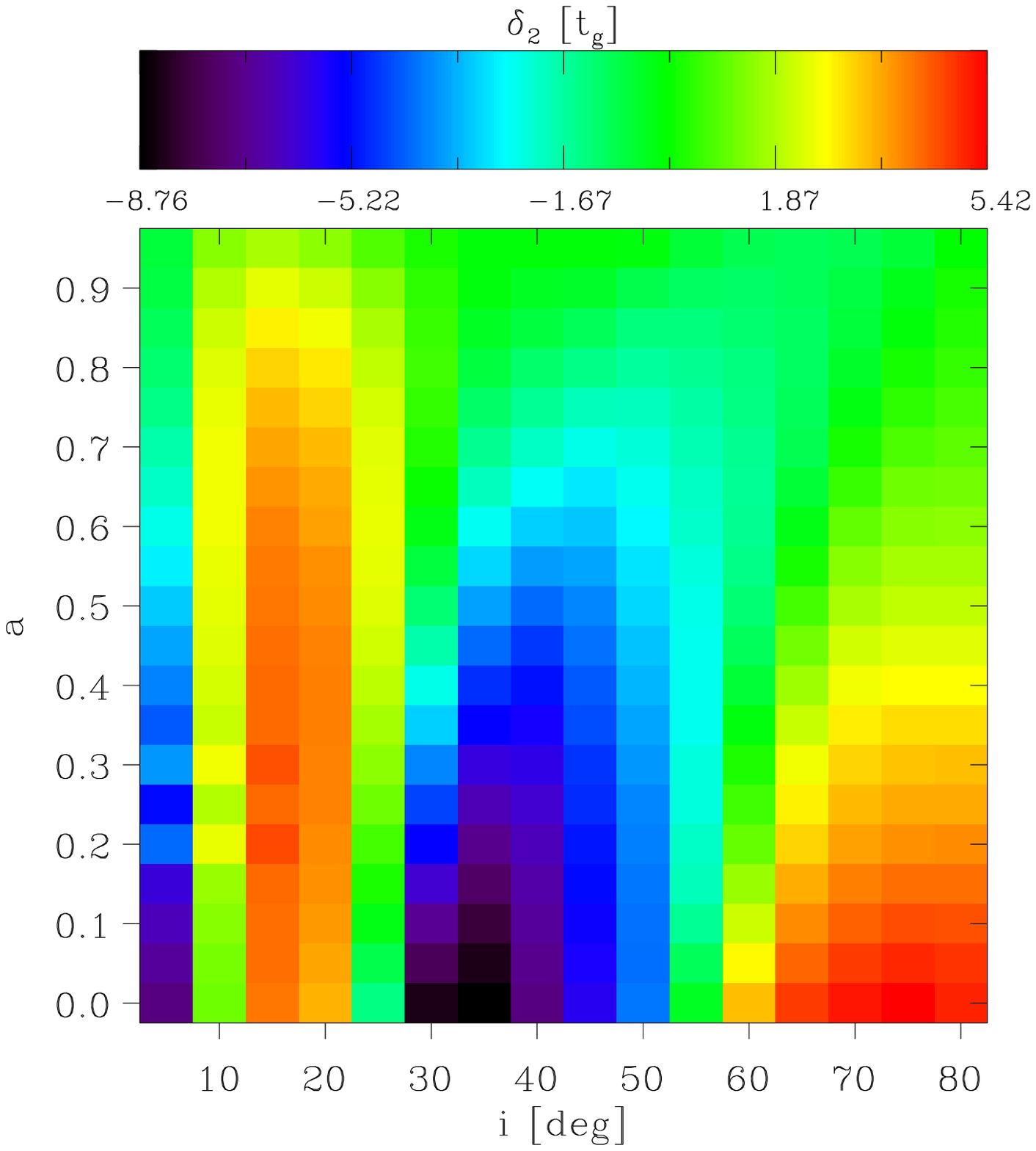}}
\centering{\includegraphics[width=0.33\textwidth]{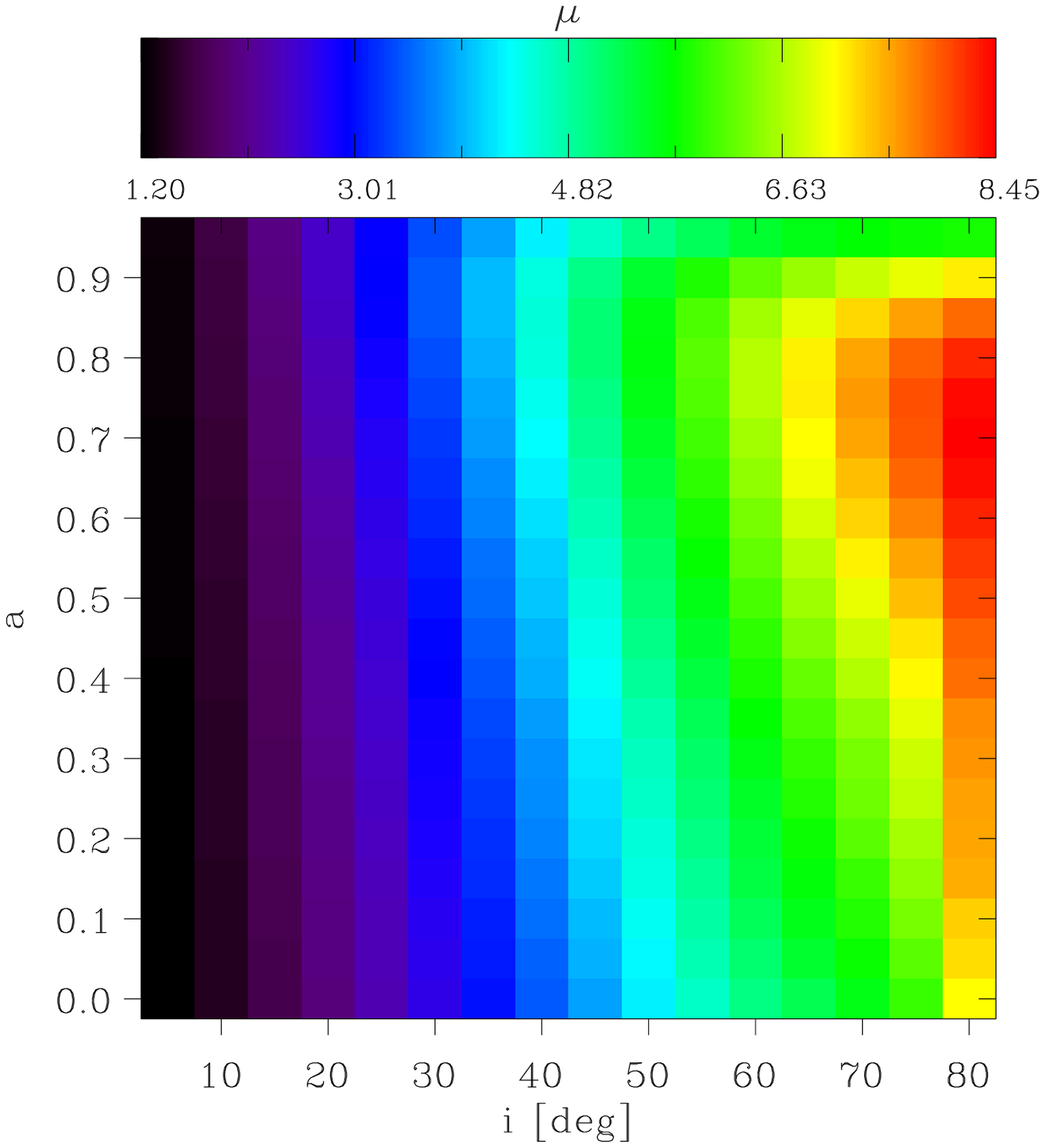}}
\centering{\includegraphics[width=0.33\textwidth]{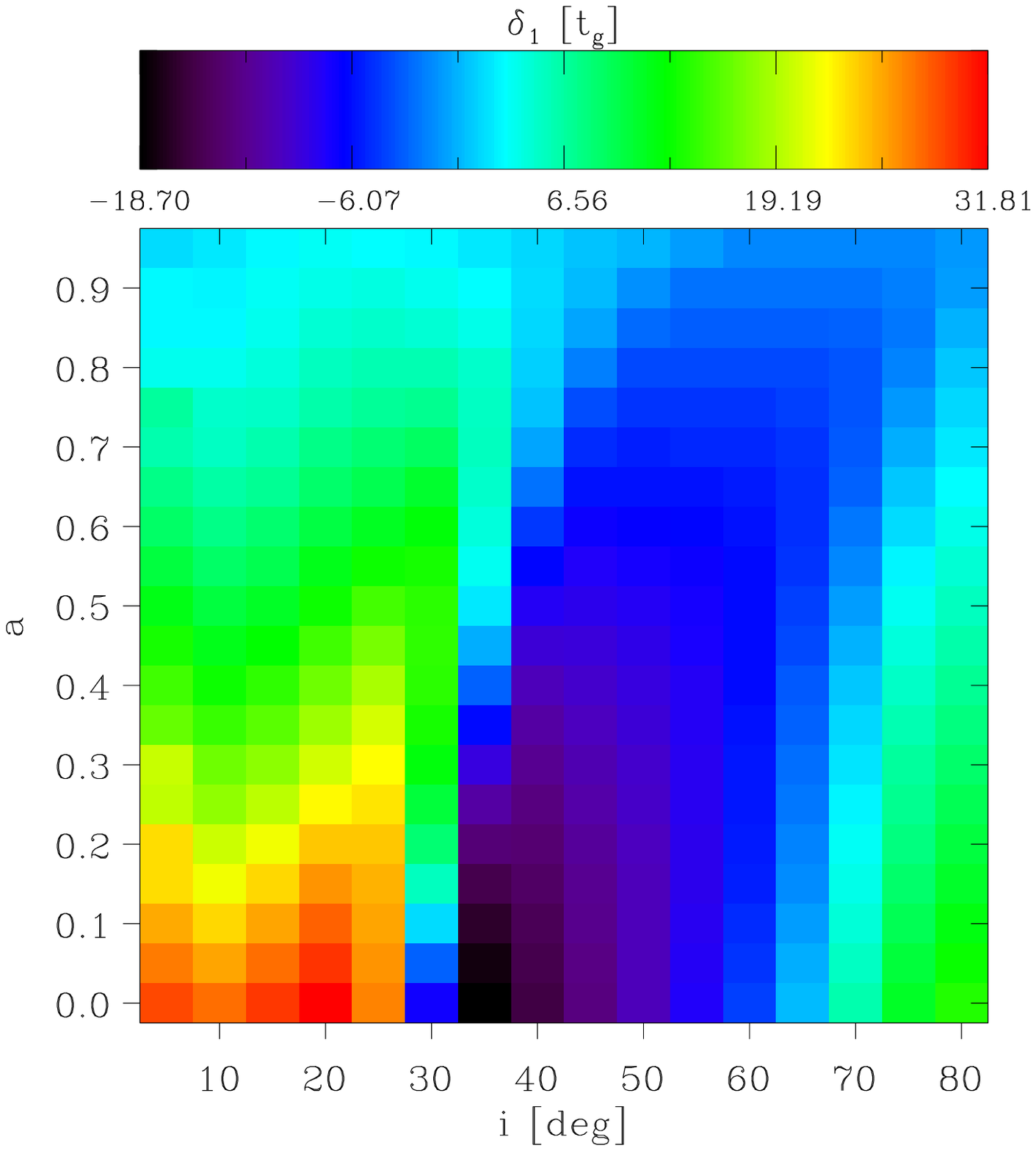}}
\centering{\includegraphics[width=0.33\textwidth]{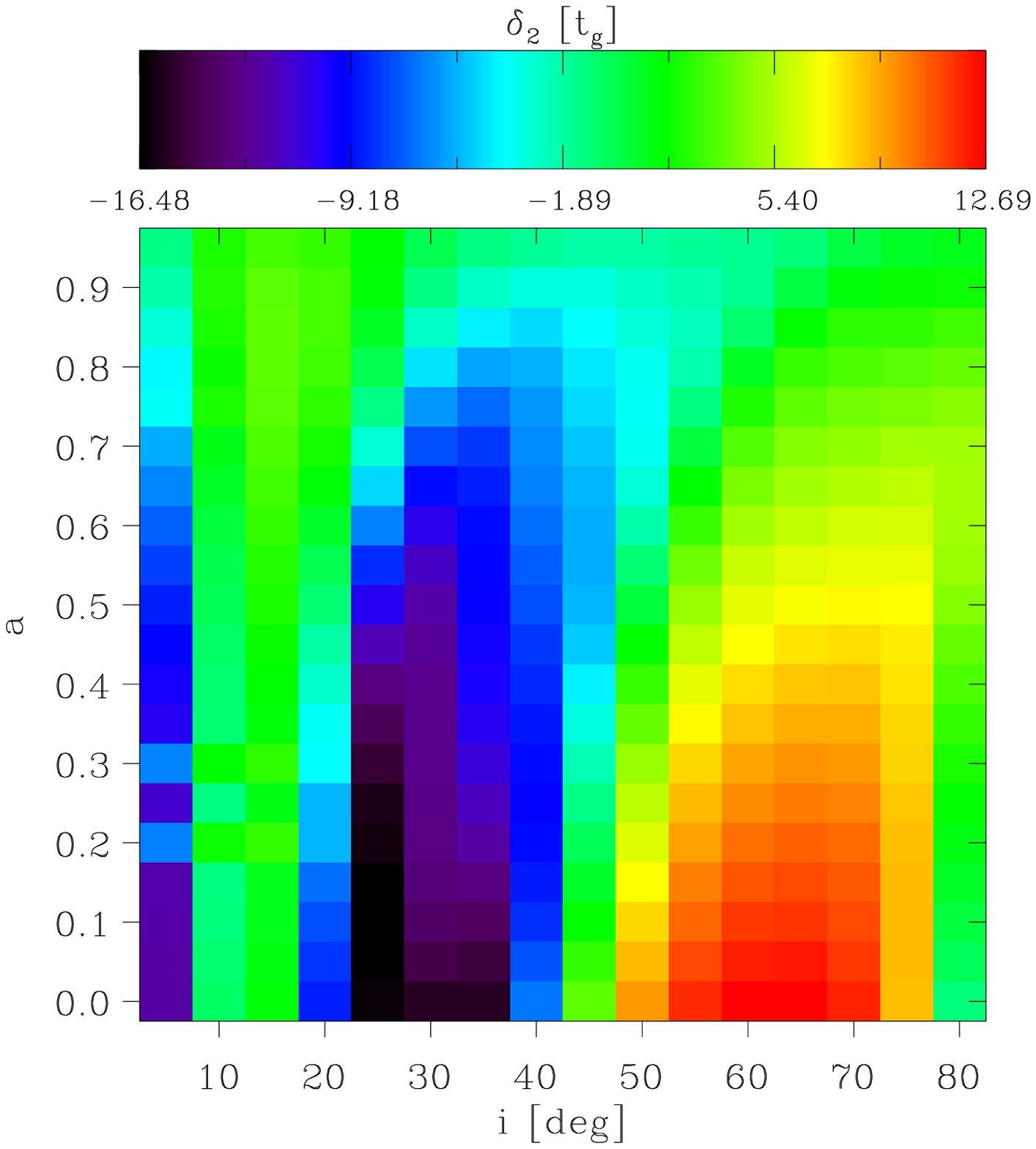}}
\centering{\includegraphics[width=0.329\textwidth]{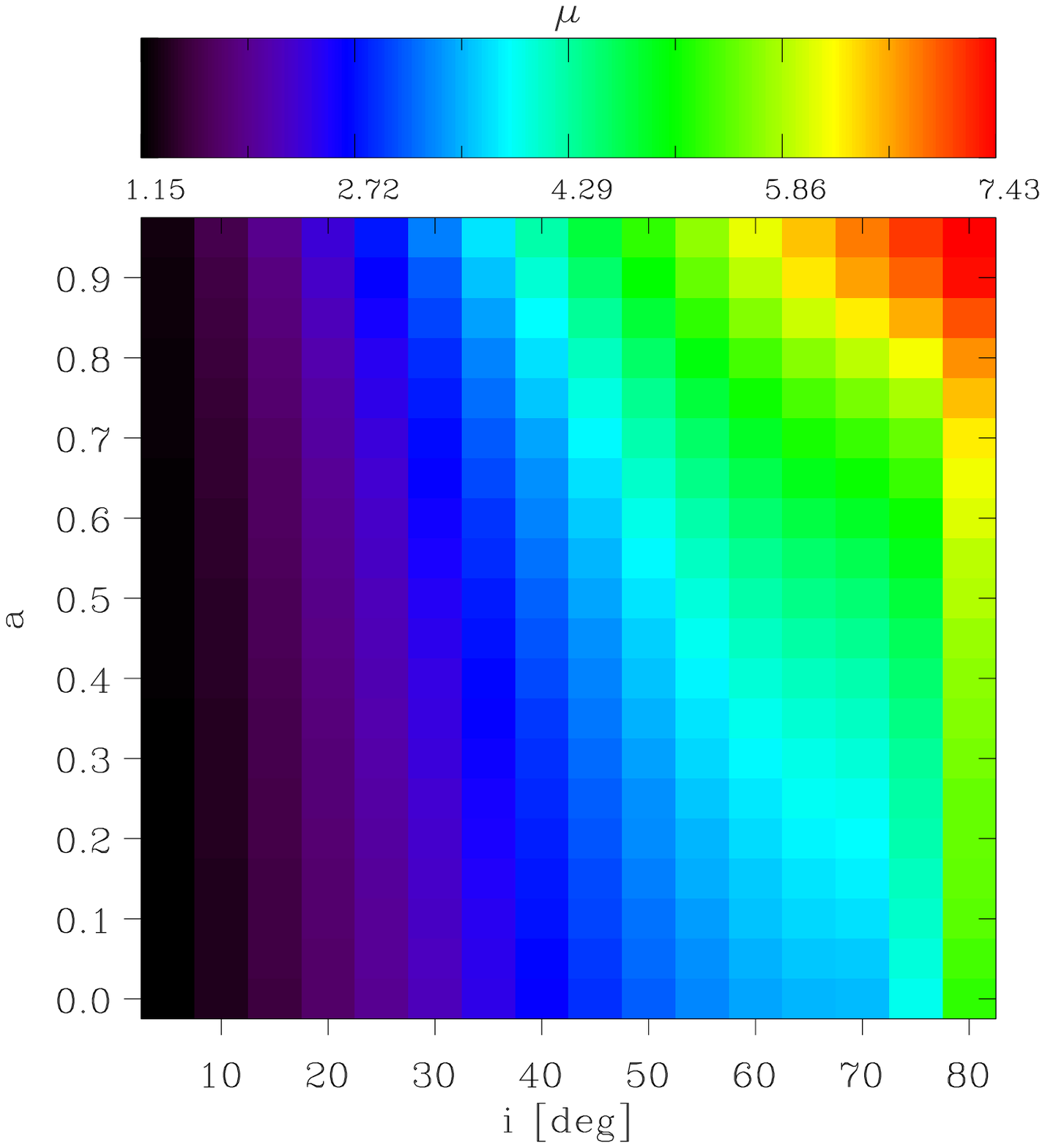}}
\caption[]{Time lag between the orthogonal polarization channels 
$\delta_1(\equiv\delta_{0-90})$, $\delta_2(\equiv\delta_{45-135})$ and flux magnification $\mu$,
as functions of  spin and inclination (from left to right, respectively).
The top row corresponds to $\xi=1.0$ while the bottom row shows the same plots for $\xi=1.5$.
Here we have assumed that the projected spin axis is aligned  toward the north on the observer's sky ($\theta=0^\circ$).
Values are presented in gravitational unit time scale ($t_g$).}
\label{complex1}
\end{figure*}

Since the majority of the high-frequency photons originate from plasma very 
close to the black hole, the curved structure of  
space-time must be taken into account. In our simulations, we have used Karas-Yaqoob (KY) code
for the ray-tracing (see Dov\v{c}iak et al. 2004 for details of the ray-tracing method).
Light bending, aberration, Doppler boosting, and frame dragging affect 
both flux and polarization properties of the emission. The KY code is able 
to calculate all the relevant relativistic effects for 
the Kerr space-time. The observed flux density from the 
accretion disc at a certain observed frequency, $F_{\nu_o}$, can be computed as follows 
(Dov\v{c}iak et. al. 2004b, 2008a,b):
\begin{equation}
F_{\nu_o}=\frac{1}{D^2}\int_{A}{G(r,\phi)I_\nu  dA}
\end{equation}
where $I_\nu$ is the intensity, $D$ is the distance to the source, $dA=rdrd\phi$ is an area element on the disk and 
$G(r,\phi)$ is the transfer function which contains all the above mentioned relativistic 
effects. This function depends on the location of the emitter
in the disc.

The hot spot is modelled as an over-density of non-thermal electrons
centred at a point orbiting at the Keplerian velocity with a Gaussian
profile as measured in the co-moving frame:
\begin{equation}
n=n_0\exp\bigl[{\frac{-(\Delta r)^2}{2 R_{spot}}}\bigl],
\end{equation}
where $\Delta r = | \overrightarrow{r} - \overrightarrow{r}_{sp}| $ with $\overrightarrow{r}_{sp}$
being the vector pointing to the centre of the spot.
We have chosen the 
typical value size of one Schwarzschild radius for the size of the spots. For the underlying steady accretion flow,
we assumed that both non-thermal electron density and the magnetic field strength scale as $r^{-1}$. Since our analysis 
is focused on high-frequency regimes (NIR and X-ray), we have ignored radiative-transfer effects. 

Here we must note that some models propose strong deviations from Keplerian velocity for the 
accretion flow. In the absence of an unambiguous theory for the accretion model, we have adopted the 
simple case of the Keplerian velocity distribution. While our choice of a thin accretion disk reduced 
the computational time needed for the ray-tracing simulations and made it possible to cover 
a wide range of free parameters in a reasonable time, it can produce large discrepancies at high inclinations.
For this reason, we have limited ourselves to inclinations less than $80^\circ$. The dimensionality 
of the spot is also important for polarized ray-tracing along strongly lensed rays. In this case, many different 
magnetic field orientations could be encountered (as seen by an observer who is parallel propagating his orientation
along the rays). For high spin values this can produce substantial depolarization. Our comparison with three-dimensional simulations with the same setup,
showed that for inclinations less than $75^\circ$ and black hole spin below $0.9$ the deviations of the models are less than $15\%$ 
while above this values it can reach as high as $40\%$.

\begin{figure}
\centering{\includegraphics[width=0.49\textwidth]{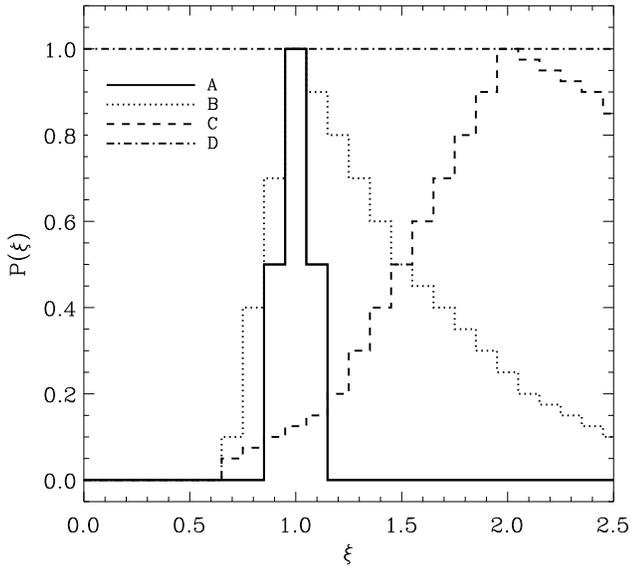}}
\caption[]{Normalized probability of  creation of spots inside the accretion disk
  as a function of distance from the black hole ($\xi$). 
  Different curves correspond to different models: (A) spots always created very close to the 
  ISCO, (B) dominant creation happens at ISCO and gradually decreases with increasing radius, 
(C) maximum probability happens at $2\times$ISCO and (D)
  equal probability everywhere inside the disk.  }
\label{models}
\end{figure}

\begin{figure}
\centering{\includegraphics[width=0.48\textwidth]{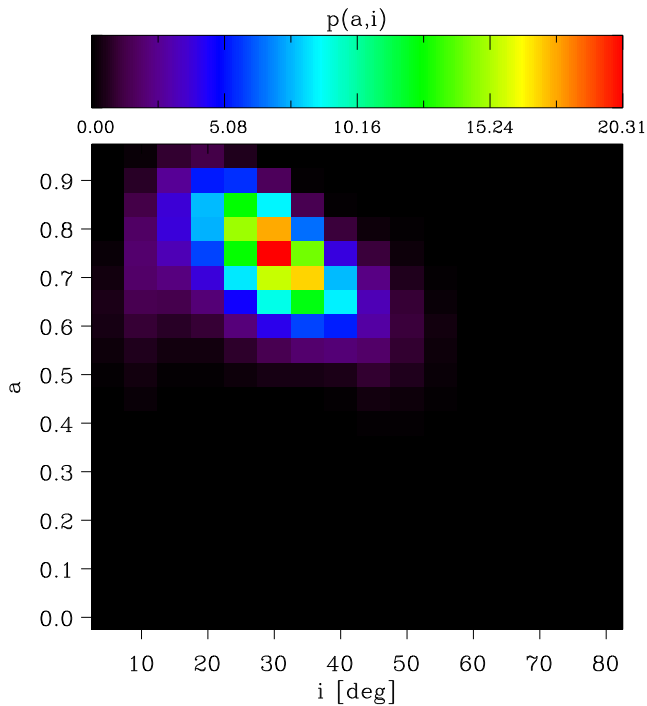}}
\caption[]{
Probability densities for a given inclination and black hole spin 
marginalised over $\theta$ for the hypothetical values of $\delta_1 = (4\pm0.5) \ t_g$,
 $\delta_2 = (3\pm0.5) \ t_g$ and $\mu = (4\pm1)$. This example shows 
how high accuracy measurements of time-lags and magnification can result in tight constraints
on $a$ and $i$.
}
\label{ai_tg}
\end{figure}

Figure \ref{image} shows the images of the assumed accretion flow at an ideal
detector as well as polarimetric images at $0^\circ$ and $90^\circ$
polarization channels for a spinning black hole ($a=0.5$) viewed by an observer inclined by $35^\circ$.
$\alpha$ and $\beta$ are the
projections of the impact parameter of the emitted photons onto the sky.
Here we have assumed that the projected spin axis of the black hole coincides with the north on the observer's sky.
We have used these images to extract light curves of the total
flux and the corresponding polarization channels.
Figure \ref{lag} shows the behaviour of flux light curves in 4 orthogonal polarization channels
(namely $0^\circ, 45^\circ, 90^\circ$ and $135^\circ$ channels)
for an orbiting blob located at the ISCO of a Schwarzschild black hole, viewed by an observer inclined by $60^\circ$.
One can see clearly that each channel reaches its maximum at a different time. 
The reason for this behaviour is the strong Doppler
beaming as well as the light focusing.
Furthermore, the rotation of the polarization plane along the photon trajectory 
plays a role. The latter effect is particularly strong for small radii of the spot
orbit (Dov\v{c}iak et al. 2008a,b).
As Fig. \ref{lag}  shows, the time-lag between each orthogonal pair of
polarization channels can be measured by cross-correlating
the corresponding light curves. One can define another parameter called \textit{magnification factor} ($\mu$) which 
provides useful information about the properties of a light curve. Here, the 
magnification factor refers to the ratio of the maximum flux density to its 
quiescent value. 
This factor is a measure 
of how much the flux is magnified by lensing and boosting effects.

Up to here, we assumed that the projected spin axis of the black hole is aligned toward the north 
on the observer's sky. In general, the measured time-delay between
two orthogonal polarization channels depends on the orientation of the projected spin axis with respect
 to the north  ($\theta$, vanishing at the north and increasing toward the east) in our chosen reference 
coordinates. In order to reduce this dependency, one can measure simultaneously time delays between two pairs
of orthogonal channels which are rotated by $45^\circ$ with relative to one another. 
Since changing the position angle
corresponds simply to a coordinate rotation, the images were originally
computed only for a single 
position angle, namely $\theta=0^\circ$. The time-lag 
associated with each desired value of $\theta$ were extracted from these images by applying the appropriate Mueller matrices.

The main question which rises in this step 
is how these time lags ($\delta_1\equiv\delta_{0-90}$ and $\delta_2\equiv\delta_{45-135}$)
 and magnification factor ($\mu$) are sensitive 
to the changes of the global parameters of our model. The main free parameters are the spin of the black 
hole ($0\leq a \leq 1$ in geometric dimensional units), inclination of the observer
 ($0^\circ\leq i \leq 90^\circ$, when the edge-on view corresponds
to $i=90^\circ$), orientation of the projected spin axis on the sky ($-90^\circ\le\theta\le90^\circ$,
 vanishing at the north and increasing toward the east), and the distance of the blob 
from the black hole (controlled by the parameter $\xi$; $r_{sp} = \xi\times \textrm{ISCO}$).
Figure~\ref{complex1} shows  
 $\delta_{1}$, $\delta_{2}$ and $\mu$ as functions of the spin and inclination for a black hole 
of the mass  $M$ in gravitational units.
The functions have been presented  for two distances of the spot from the black hole ($\xi=1.0$ and $\xi=1.5$)
 while $\theta=0^\circ$.
We have calculated these functions for a wide range of acceptable values of this parameter
 ($\xi = 0.5 - 3.0$; with a step
of 0.1). One can see that these three different functions behave differently
 in the $a-i$ space. 
This allows us to use these functions
 as tools for constraining spin, inclination and orientation of the accreting systems. 

\begin{figure*}
\centering{\includegraphics[width=0.33\textwidth]{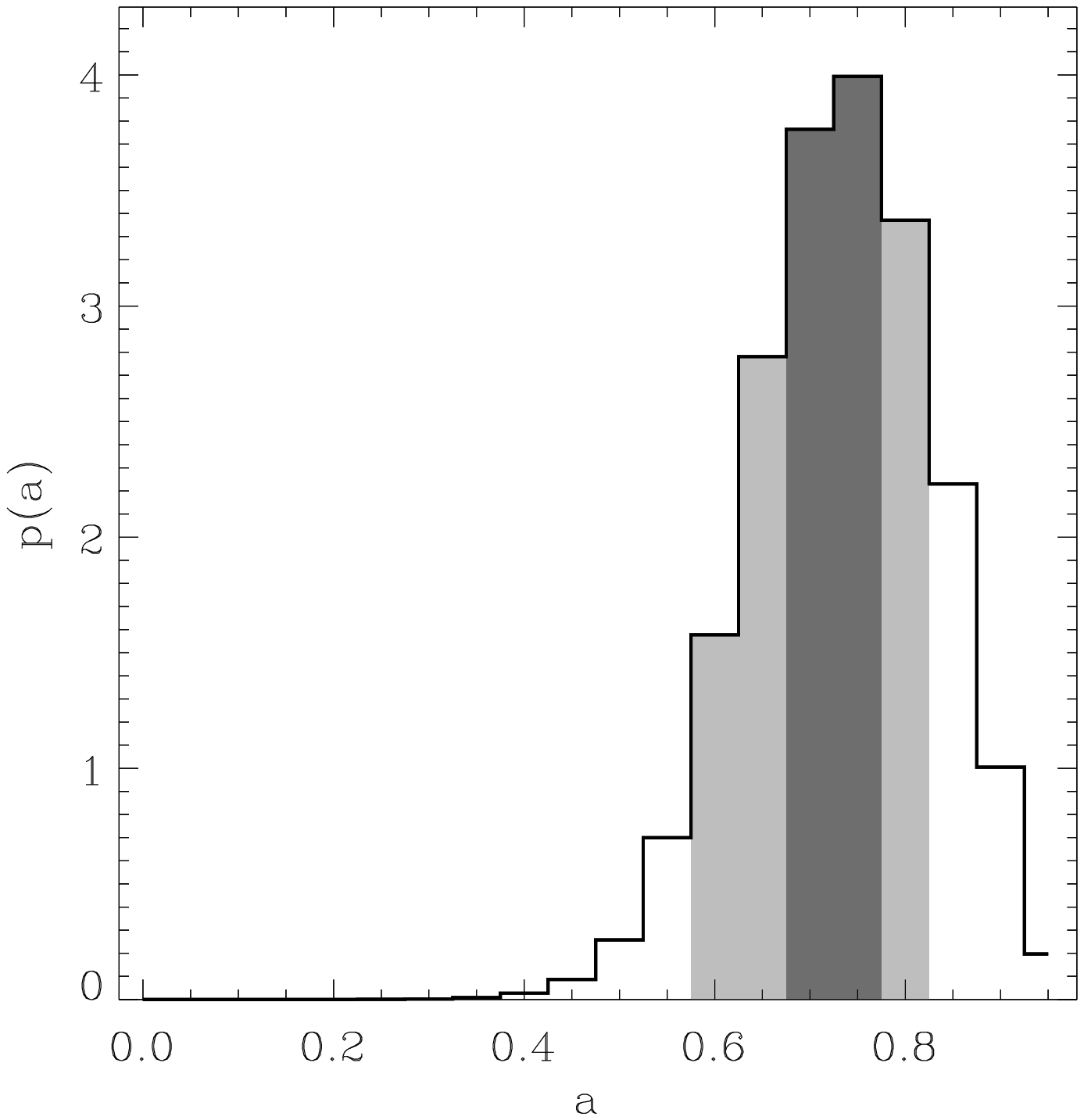}}
\centering{\includegraphics[width=0.33\textwidth]{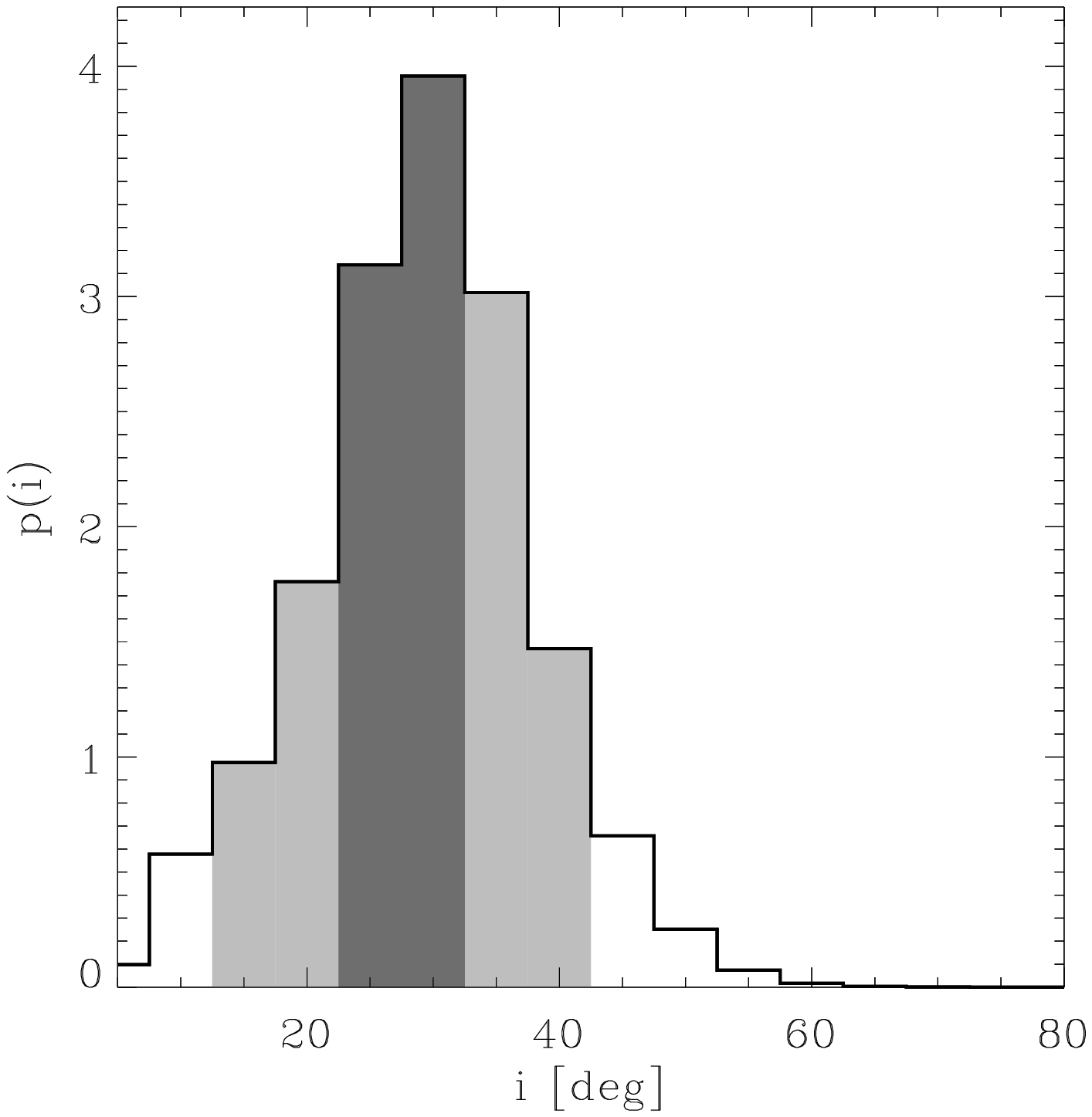}}
\centering{\includegraphics[width=0.33\textwidth]{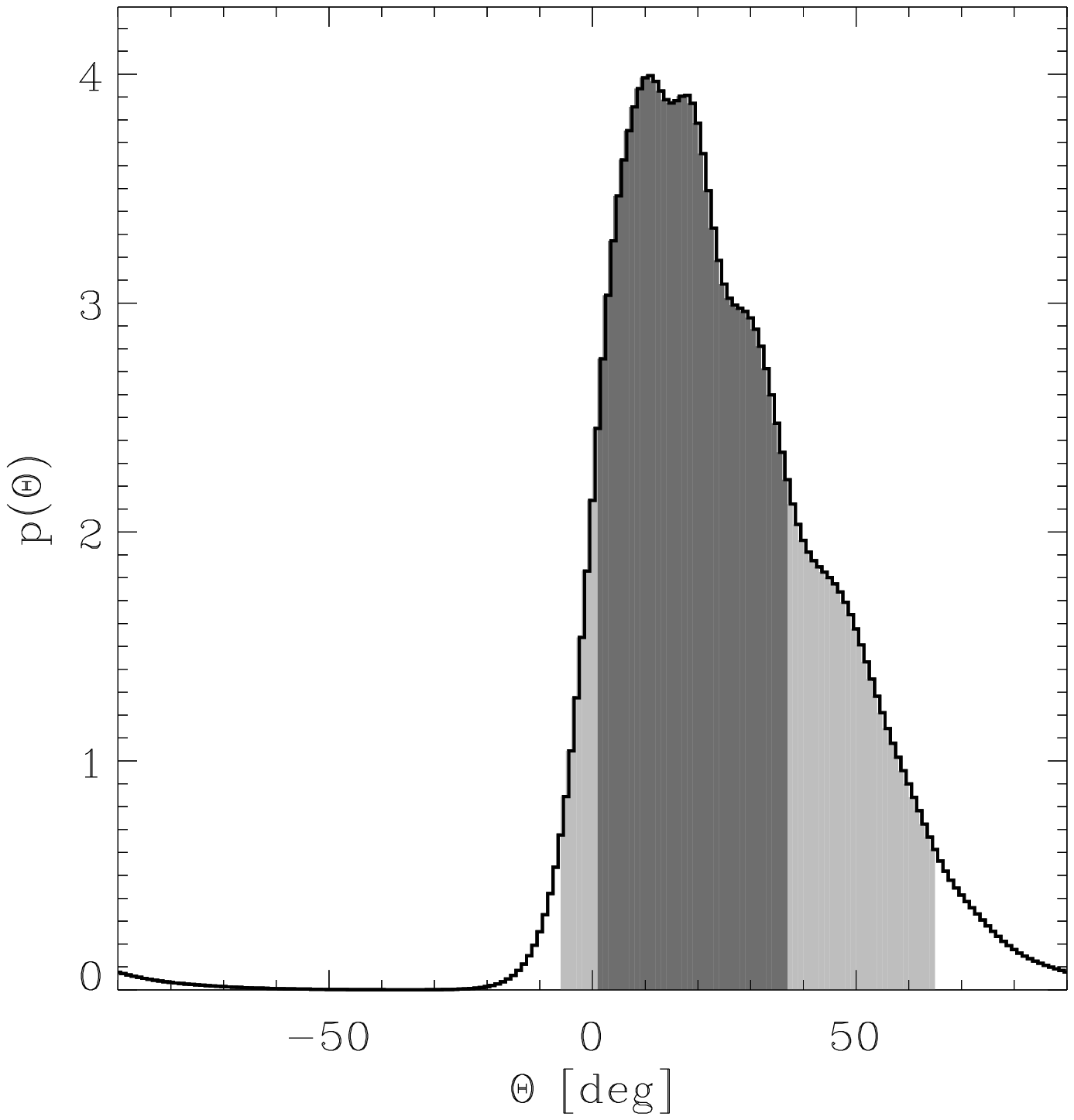}}
\caption[]{
Probability densities for a given spin $p(a)$, inclination $p(i)$ and 
orientation of the system on the sky $p(\theta)$,  marginalised over other parameters (left to right, respectively).
The assumed values of $\delta_1$, $\delta_2$ and $\mu$ are the 
same as in Fig.~\ref{ai_tg}.
The dark and light shaded areas denote the $1\sigma$ and $2\sigma$ confidence limits, as defined in the 
text, respectively. }
\label{ait_tg}
\end{figure*}

It is clear from  Fig. \ref{complex1} that our results are sensitive to the exact location of the orbiting
blob with respect to the black hole (see
predictions for $\xi = 1.0$  and $\xi = 1.5$ in Fig.~\ref{complex1}). Unfortunately, there is no direct way 
to constrain that parameter. A lot of effort have been devoted to understand where the radiation and the stress edges
of an accretion disk are located and how they depend on the different assumptions of the models
 (Beckwith et al. 2008, Shafee et al. 2008, Hilburn et al. 2010, Noble et al. 2009, 2010).
Since it is not yet clear how and where  spots can be created inside an accretion disk, 
we have assumed four different types of models for the spatial distribution of the spots 
as a function of distance from the black hole (Fig.~\ref{models}). We have considered models in which 
the main creation of spots happens only around the ISCO (model A), distributed over the disk with the maximum 
at the ISCO and gradually decreases with increasing the radius (model B), distributed over the disk with the maximum 
at the $2\times$ ISCO (model C, see Shafee et al. 2008) and an equal distribution
for the probability of a spot being located anywhere inside the disk (model D).
We have used the product of the magnetic stress and energy 
dissipation functions from MHD simulations (e.g. Hilburn et al. 2009 and Shafee et al. 2008)
in order to define such radial distributions. We have adopted a Bayesian  parameter estimation 
method used by Broderick et al. 2009. Here, we repeat the basics of the method  in order 
to keep the paper self-contained (see Broderick et al. 2009 for more details about the method).

The constraints upon $a$, $i$ and $\theta$ we report here, suffer from the uncertainty introduced by 
the assumption that our model is appropriate for the observations. We compute the probability 
following a Bayesian scheme, which states that a given set of model parameters
($a$,$i$,$\theta$) are correct for the measured delays and magnifications.

Given a particular set of model parameters $(a,i,\theta,\xi)$, the probability of 
measuring  an observable $x_j$ (where $x_j\in[\delta_{1j},\delta_{2j},\mu_j]$), 
assuming Gaussian observational errors, is
\begin{equation}
\begin{array}{l}
P_j(x_j|a,i,\theta,\xi)
= \nonumber\\
\displaystyle
\quad\frac{1}{\sqrt{2\pi}\Delta x_j}
\exp\left\{
- \frac{\left[x_j - x(a,i,\theta,\xi)\right]^2}{2\Delta x_j^2}
\right\}
d x_j
\,.
\end{array}
\end{equation}
Therefore, the probability of simultaneously observing a measured set of
 $[\delta_{1j},\delta_{2j},\mu_j]$ for a certain model is
\begin{equation}
\begin{array}{l}
\displaystyle
P(\{\delta_{1j},\delta_{2j},\mu_j\}|a,i,\theta,\xi)= \nonumber\\
P(\delta_{1j}|a,i,\theta,\xi) \times P(\delta_{2j}|a,i,\theta,\xi) \times P(\mu_{j}|a,i,\theta,\xi).
\label{eq:Pvp}
\end{array}
\end{equation}
Contrary to  eq. (\ref{eq:Pvp}), which gives the probability that the observed quantities come 
from a certain model, choosing an appropriate set of priors on $a$, $i$, $\theta$ and $\xi$ allows 
us to compute, via Bayes' theorem, the probability density of a set of model parameters given 
the observed values, i.e. $p(a,i,\theta,\xi|\{x_i\})$.  As such, we now turn to the
problem of choosing these priors.

Assuming the isotropic probability of the black hole's spin orientation, a choice for the 
prior upon $i$ and $\theta$ can be made so that  $\wp(i,\theta) = \sin(i)$ (Broderick et al. 2009). 
The prior on $a$ is chosen to be uniform, i.e. $\wp(a)=1$ since we lack a complete theoretical 
knowledge about the spin evolution of super-massive black holes. Finally, 
the prior on $\xi$ is set according to the models in Figure~\ref{models}. Therefore, Bayes' theorem gives
\begin{equation}
\begin{array}{l}
\displaystyle
p(a,i,\theta,\xi|\{x_j\})
\nonumber\\
\\
\qquad\displaystyle
=
\frac{\displaystyle
P(\{x_j\}|a,i,\theta,\xi) \wp(a) \wp(i,\theta)\wp(\xi)
}{\displaystyle
\int da \ di  \ d\theta \ d\xi \, P(\{x_j\}|a,i,\theta,\xi) \wp(a) \wp(i,\theta) \wp(\xi)
}
\nonumber\\
\\
\qquad\displaystyle
=
\frac{\displaystyle
P(\{x_j\}|a,i,\theta,\xi) \sin(i)
}{\displaystyle
\int\ da  \ di \ d\theta \ d\xi \, P(\{x_j\}|a,i,\theta,\xi) \sin(i)
}\,.
\end{array}
\end{equation}

As it can be seen, this is a probability density in a four-dimensional parameter space, 
which makes it difficult to be visualized directly.
Therefore, we construct a variety of marginalised probabilities for presentation and analysis. 
This is done by forming a pair of two-dimensional $p(a,i)$ function with the probability densities marginalised over $\theta$ and $\xi$:
\begin{equation}
p(a,i) = \int d\theta \ d\xi \, p(a,i,\theta,\xi|\{x_j\})
\label{eq:marg_xi}
\end{equation}
This is plotted in Fig.~\ref{ai_tg} for a hypothetical example of a source 
with the measured delays of $\delta_1 = (4\pm0.5)\ t_g$ (where $t_g$ is the 
gravitational time) and $\delta_2 = (3\pm0.5)\ t_g$ while the flux is magnified 
by a factor of $\mu=(4\pm1)$. In an observationally-constrained astrophysical case, 
it is difficult to determine the values of the time delays and magnification 
so precisely. As one can see, the model allows to exclude a wide range of possible $a$-$i$ 
combinations and well constrains the possible range of inclination and black hole's spin. 
In order to distinguish the probability distribution of each parameter, the marginalised 
one-dimensional probability densities are also constructed:
\begin{eqnarray}
p(a) &= \displaystyle \int d\theta \ di \ d\xi \, p(a,i,\theta,\xi|\{x_j\})\\
p(i) &= \displaystyle \int d\theta \ da \ d\xi \, p(a,i,\theta,\xi|\{x_j\})\\
p(\theta) &= \displaystyle \int di \ da \ d\xi   \, p(a,i,\theta,\xi|\{x_j\})\,.
\label{eq:marg1_a}
\end{eqnarray}
These are shown in Fig.~\ref{ai_tg} for the same hypothetical values of $\delta_1$, $\delta_2$ 
and $\mu$ as in Fig.~\ref{ait_tg}. For both Figs.~\ref{ai_tg} and \ref{ait_tg} we have used the 
commonly used assumption that the spot is located at the ISCO (model A for $p(\xi)$). It is evident that the allowed parameter 
space is highly non-Gaussian.  As a result, the extraction of values and their associated uncertainties for the 
fitting parameters must be done carefully. These values will be highly correlated in all occurrences, 
and the systematic uncertainties due to assuming a particular accretion model will dominate the errors (and thus will not be reviewed again here).

For determining the $1\sigma$ and $2\sigma$ error intervals, we used an additive probability as defined in Broderick et al. 2009:
\begin{equation}
P(>p) = \int_{p(\bmath{x})\ge p} p(\bmath{x}) \, d\bmath{x}
\end{equation}
where $\bmath{x}$ is the parameter of the total probability distribution. The above definition of $P(>p)$ gives 
the probability associated with the parameter space region that has a probability density above $p$. We set 
the $1$-$\sigma$ and $2\sigma$ contours to indicate the regions of $p$ for which $P(>p)=0.683$ and $P(>p)=0.954$,
 respectively. These follow the normal definition of $1\sigma$ and $2\sigma$ errors, while in our case the errors 
are strongly non-Gaussian. In Fig.~\ref{ait_tg} the $1\sigma$ and $2\sigma$ confidence intervals are indicated by the dark and light shaded regions.

In the next section, we  will  discuss the implications of our model for the observed variable NIR polarized emission from Sagittarius~A*.

\section{Observations}
\label{sgra_sect}
\subsection{Near-Infrared Flares of Sagittarius~A*}

\begin{figure}
\centering{\includegraphics[width=0.48\textwidth]{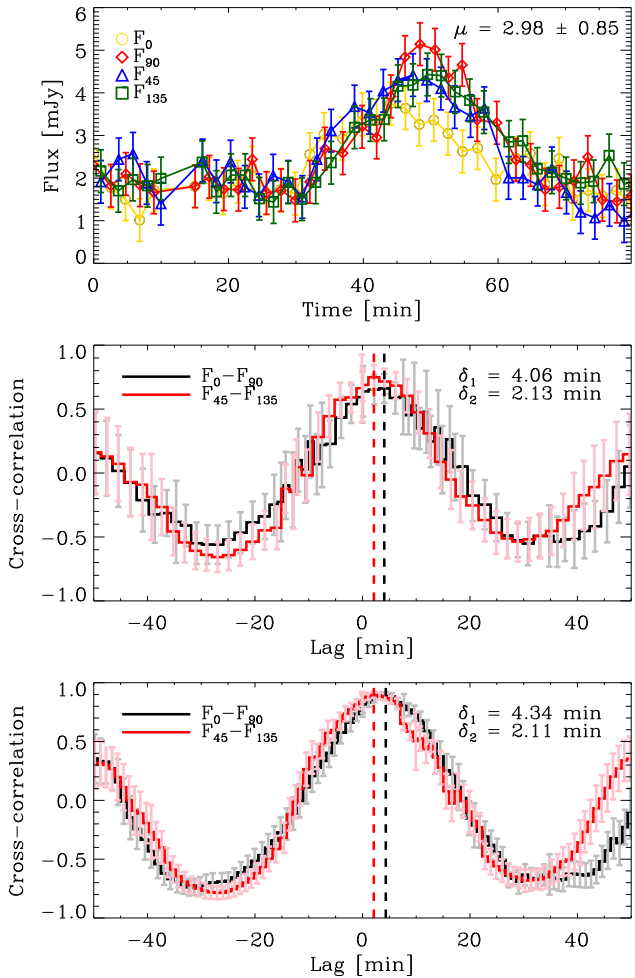}}
\caption[]{The NIR flare event of Sgr~A* observed on 13 June 2004. 
Four different polarized channels are indicated with different colours (top). 
In order to derive the time-lags a cross-correlation analysis has been performed. 
Maximum of two cross-correlation curves have been indicated
with two different 
methods (middle and bottom). 
The average  values of 
$\bar{\delta}_1=(4.20\pm1.00) \  \textrm{min}, \bar{\delta}_2=(2.12\pm1.00) \ \textrm{min}$ 
and  $\bar{\mu}=(2.98\pm1.00)$ derived 
from both analysis. 
}
\label{lag_obs1}
\end{figure}
\begin{figure*}
\centering{\includegraphics[width=0.45\textwidth]{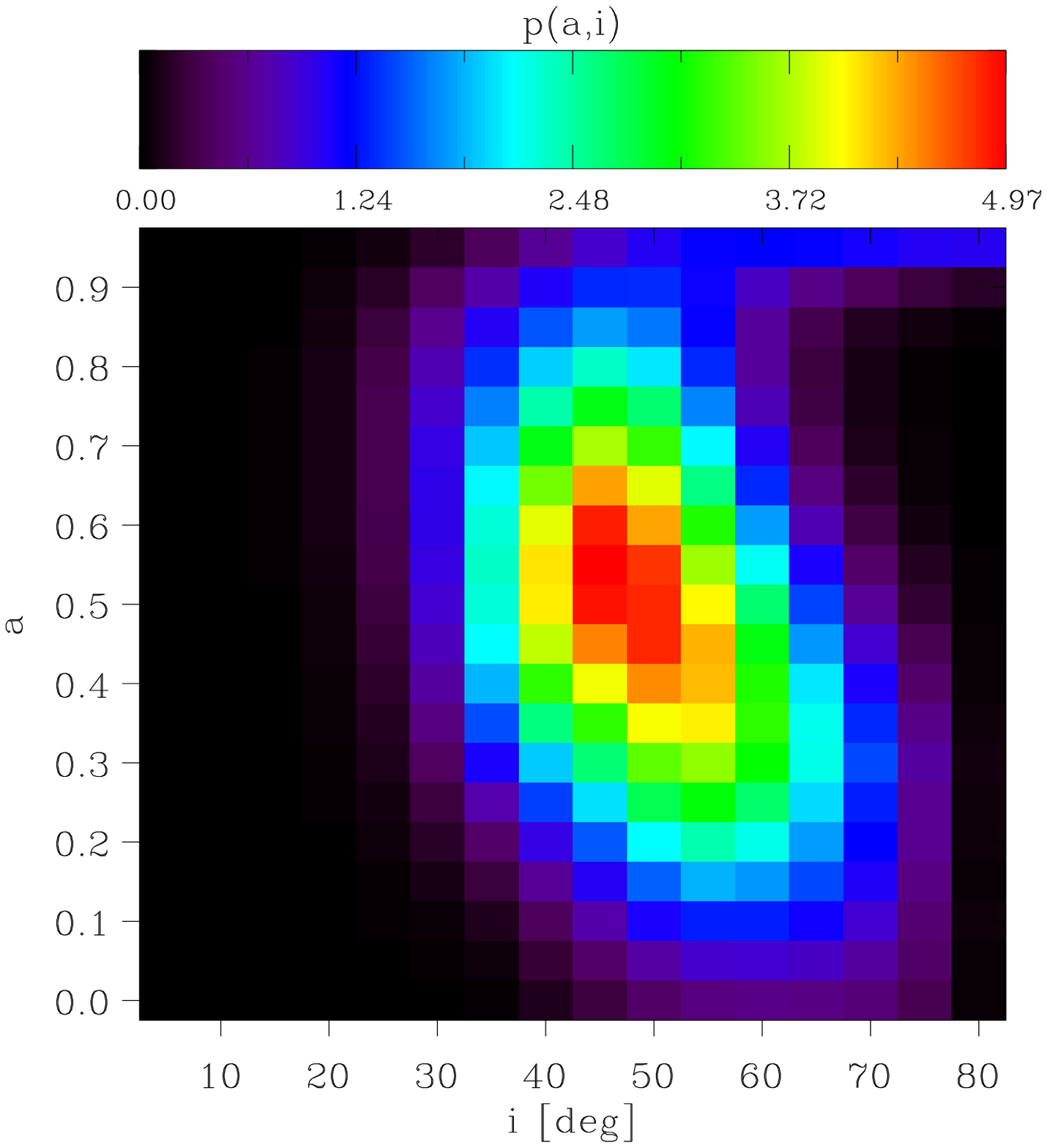}}
\centering{\includegraphics[width=0.45\textwidth]{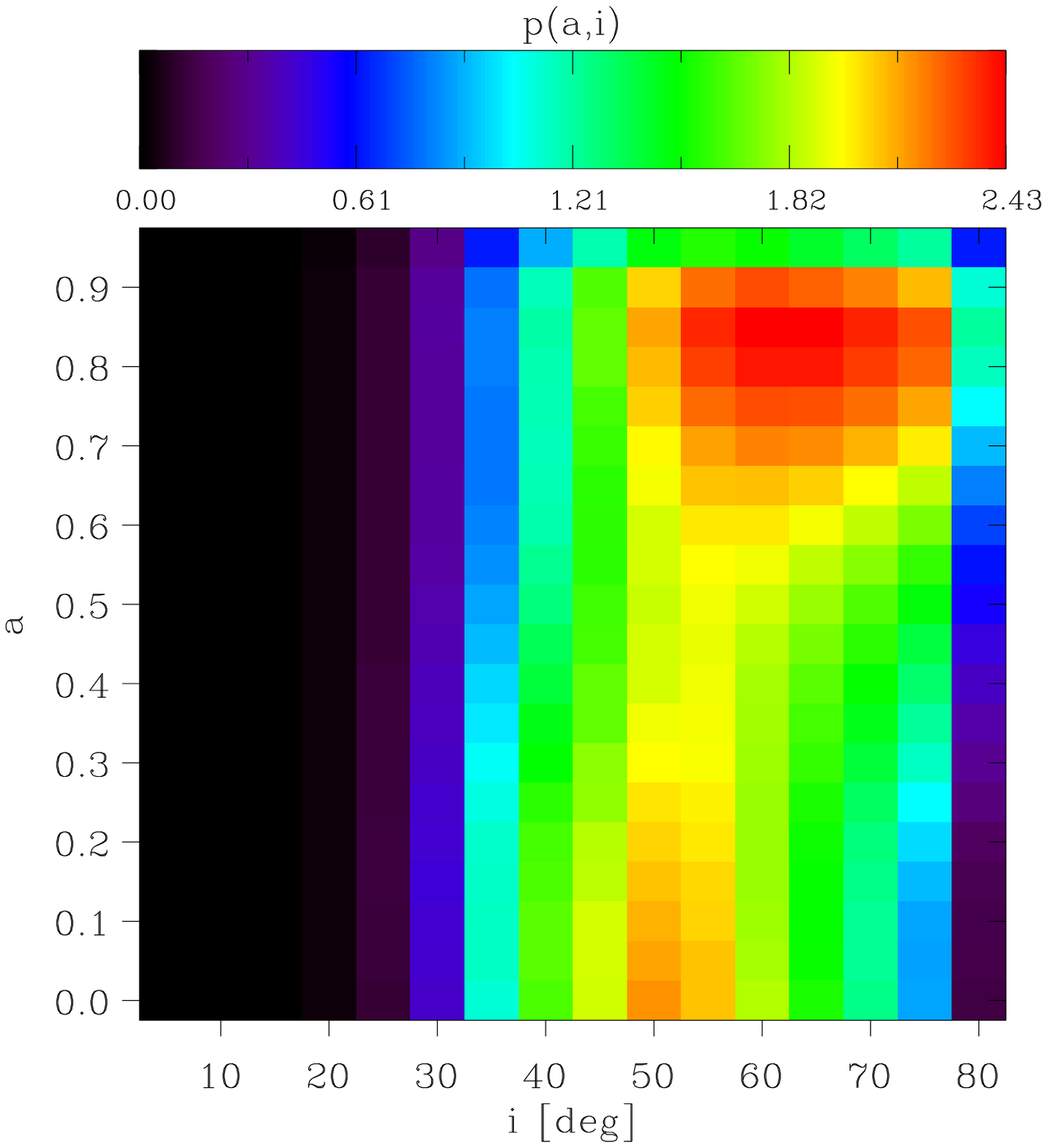}}
\centering{\includegraphics[width=0.45\textwidth]{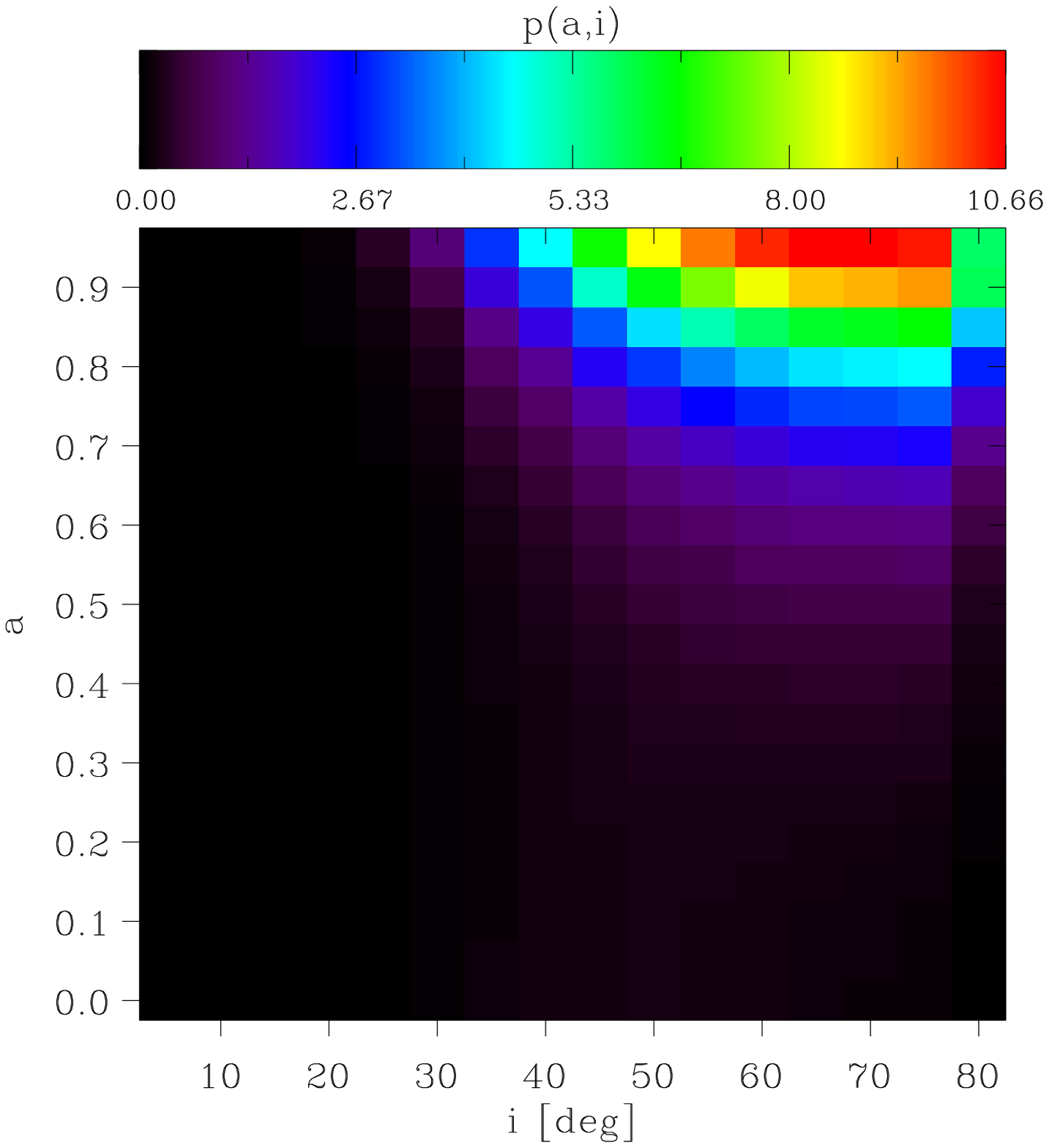}}
\centering{\includegraphics[width=0.45\textwidth]{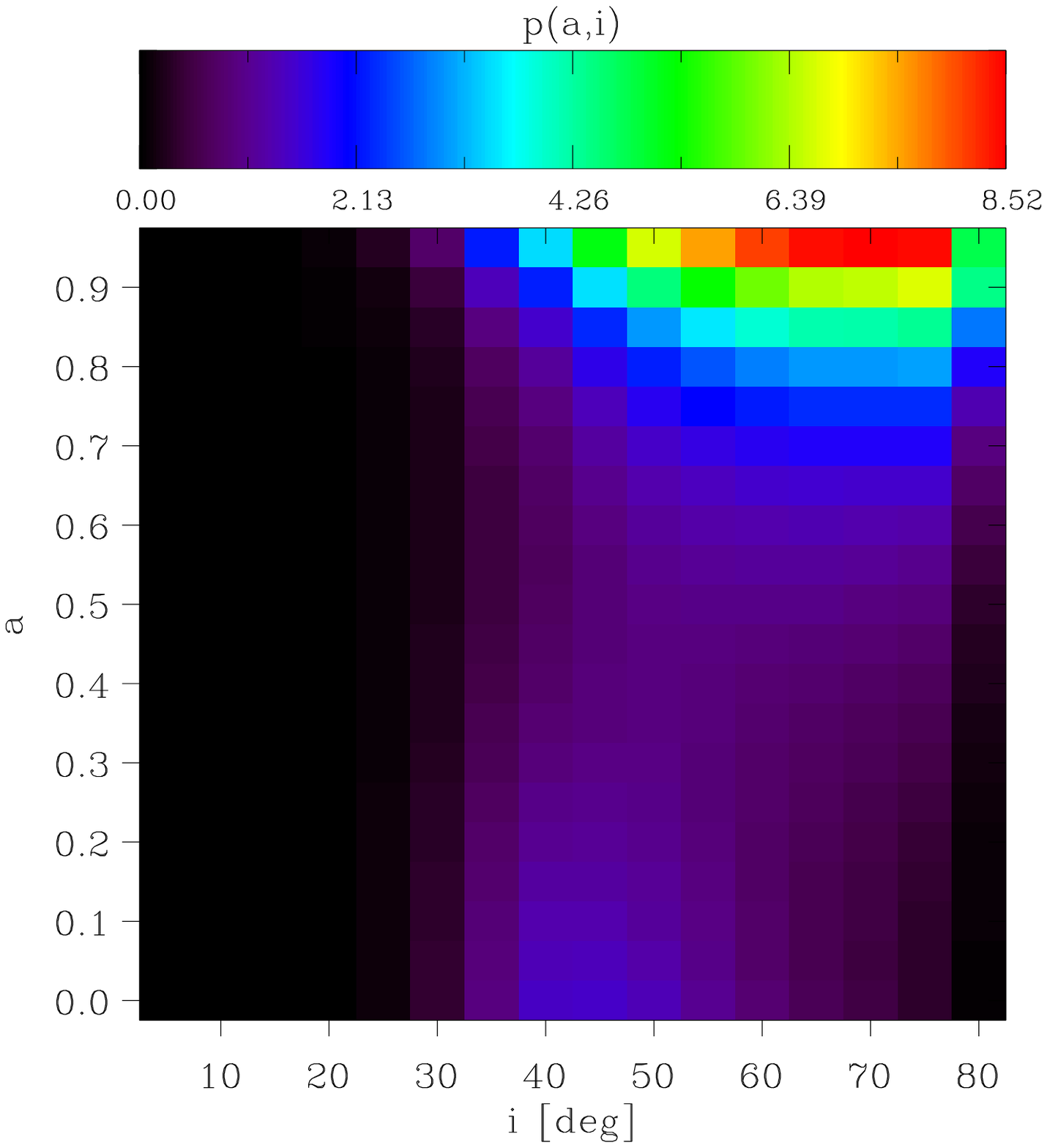}}
\caption[]{
Same as Fig. \ref{ai_tg} for the time-lag and magnification values derived from
the observed flare of Sgr~A* on 13 June 2004 (Fig. \ref{lag_obs1}). The results are shown for different models of $p(\xi)$: 
model A (top left), model B (top right),  model C (bottom left) and 
model D (bottom right).}
\label{ai_sgr}
\end{figure*}

At the centre of the Milky Way 
stellar motions allow us to firmly 
associate Sagittarius A* 
(Sgr~A*) with a $\sim4\times$10$^6$~\solm\ 
super-massive black hole 
(Eckart \& Genzel 1996, Genzel et al. 1997, 2000, Ghez et al. 1998, 2000, 2003, 2005a,2008, 
Eckart et al. 2002, Sch\"odel et al. 2002, 2003, 2009, Eisenhauer et al. 2003, 2005, Gillessen et al. 2009).

Recent radio, near-infrared and X-ray observations have detected variable 
and polarized emission and give detailed insight into the 
physical emission mechanisms at work in Sgr~A* (e.g. Baganoff et al. 2001, 2002, 2003, 
Eckart et al. 2002, 2004, 2006, 2008, 2009,
Porquet et al. 2003, 2009, Goldwurm et al. 2003, Genzel et al. 2003, 
Reid et al. 2004, Ghez et al. 2004ab, Eisenhauer et al. 2005, 
Belanger et al. 2006, Hornstein et al. 2007,
Yusef-Zadeh et al. 2006ab, 2007, 2008, 2009, Marrone et al. 2008,  
 Sabha et al. 2010).
Sgr~A* - in terms of Eddington luminosity - is the faintest super-massive black hole
known. However, due to its proximity it is bright enough to be studied in great detail. 
With the possible exception of the closest galaxies,
no extragalactic super-massive black hole with a similar feeble Eddington rate would
be observable. 

\begin{figure*}
\centering{\includegraphics[width=0.33\textwidth]{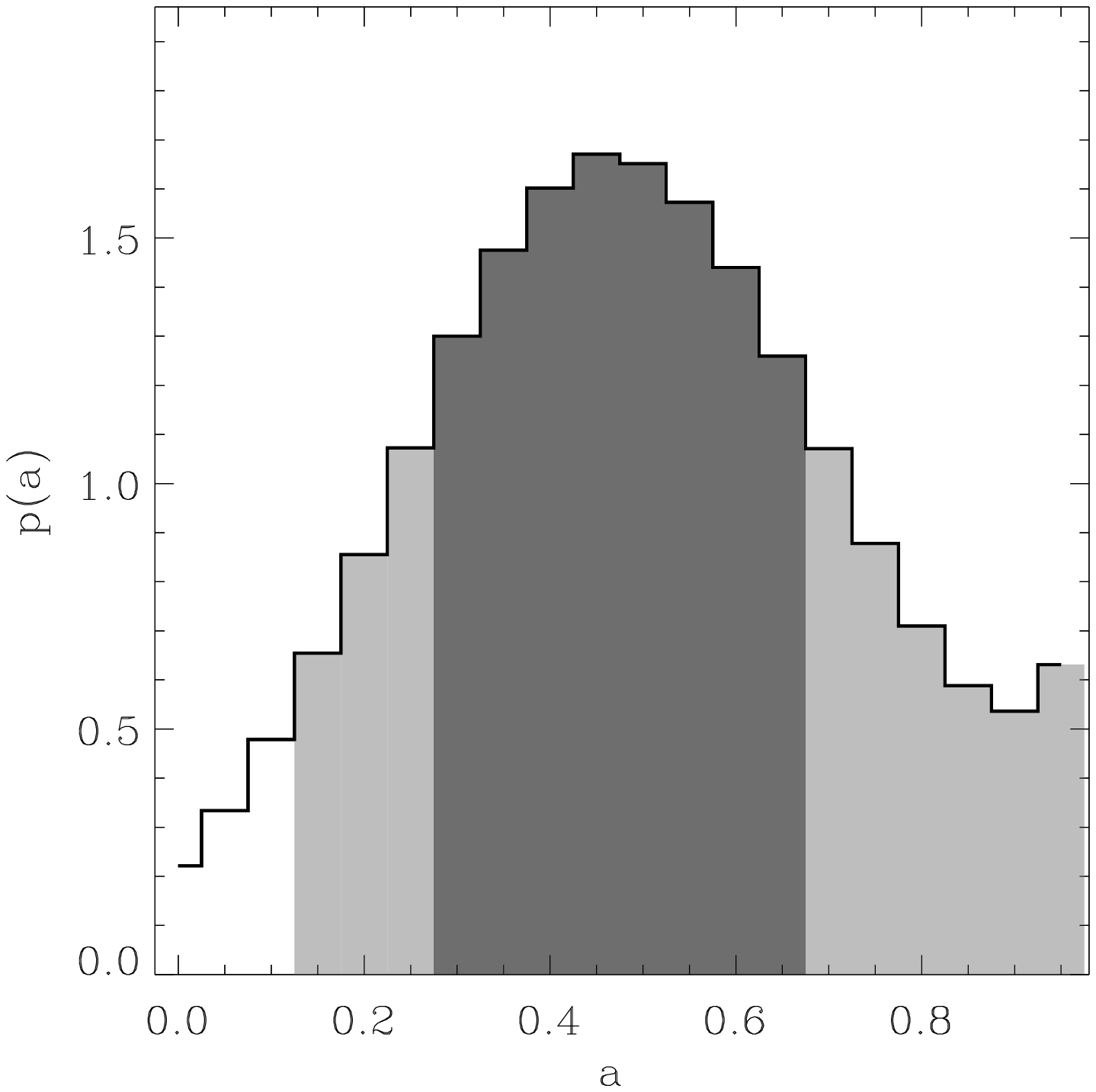}}
\centering{\includegraphics[width=0.33\textwidth]{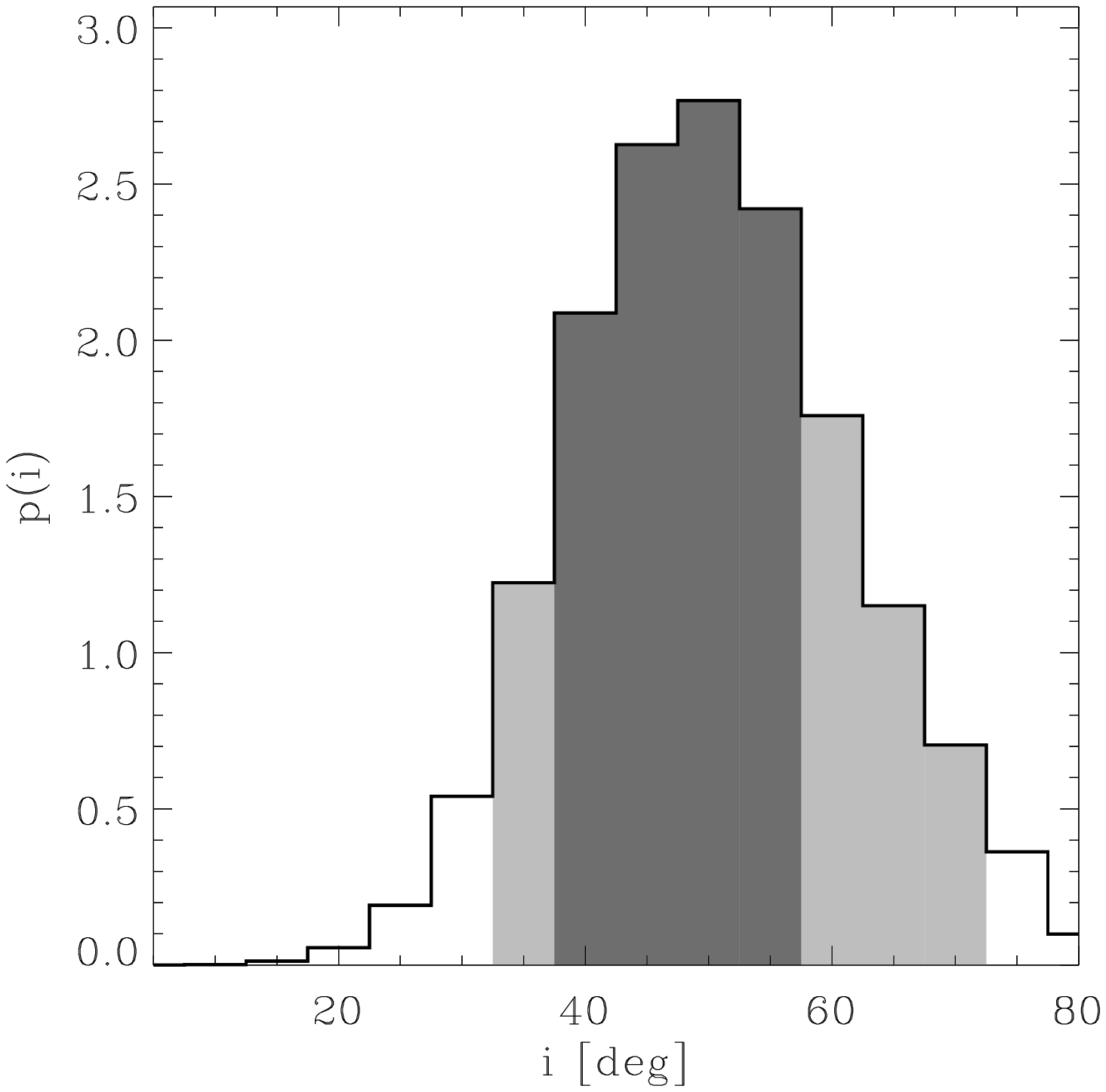}}
\centering{\includegraphics[width=0.33\textwidth]{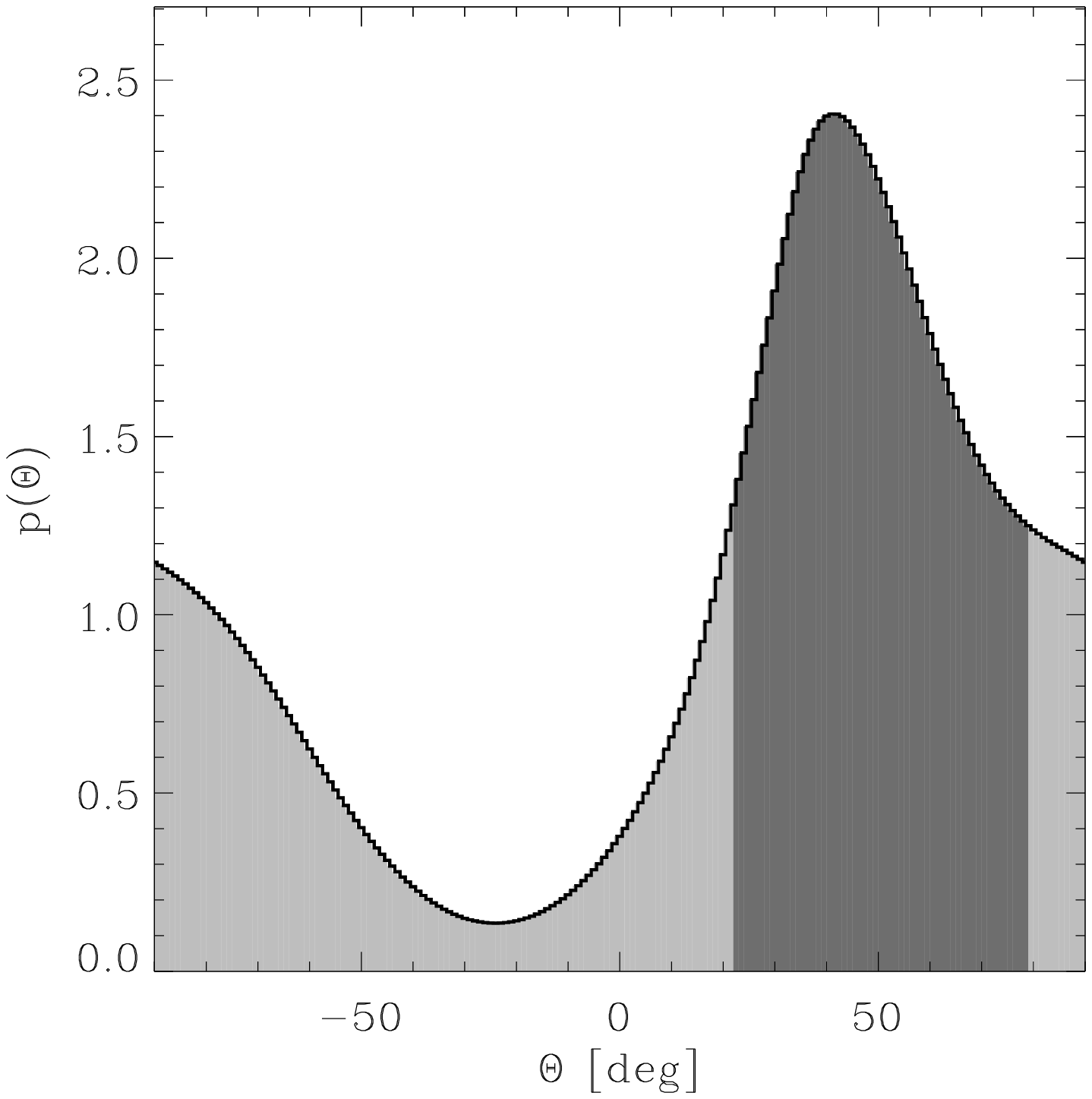}}
\caption[]{
Same as Fig. \ref{ait_tg} for the time-lag and magnification values derived from
the observed flare of Sgr~A* on 13 June 2004 (Fig. \ref{lag_obs1}). $p(a)$, $p(i)$ and $p(\theta)$ are marginalised over all 
other parameters assuming model A for $p(\xi)$.
}
\label{modelA}
\end{figure*}
\begin{figure*}
\centering{\includegraphics[width=0.33\textwidth]{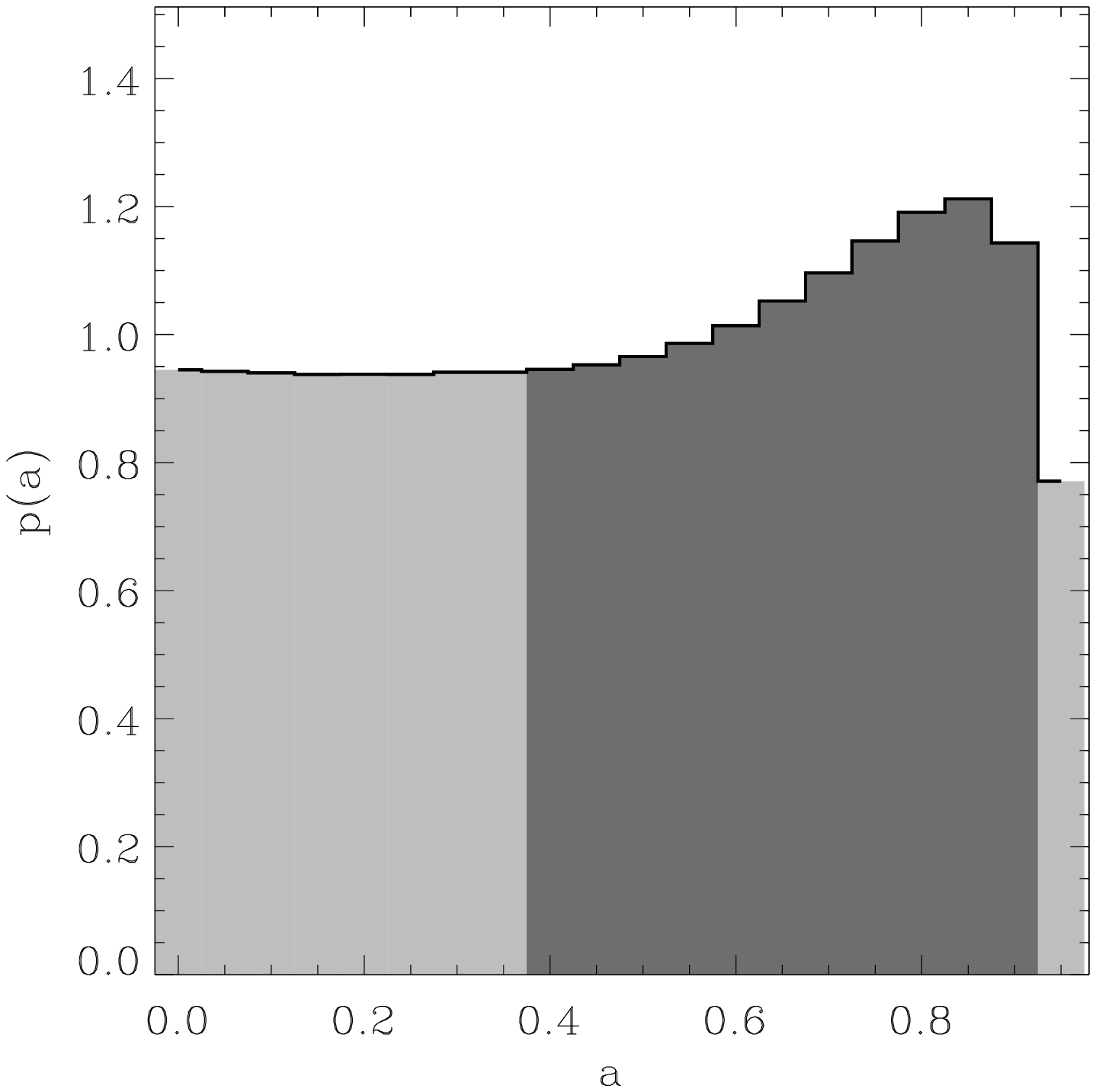}}
\centering{\includegraphics[width=0.33\textwidth]{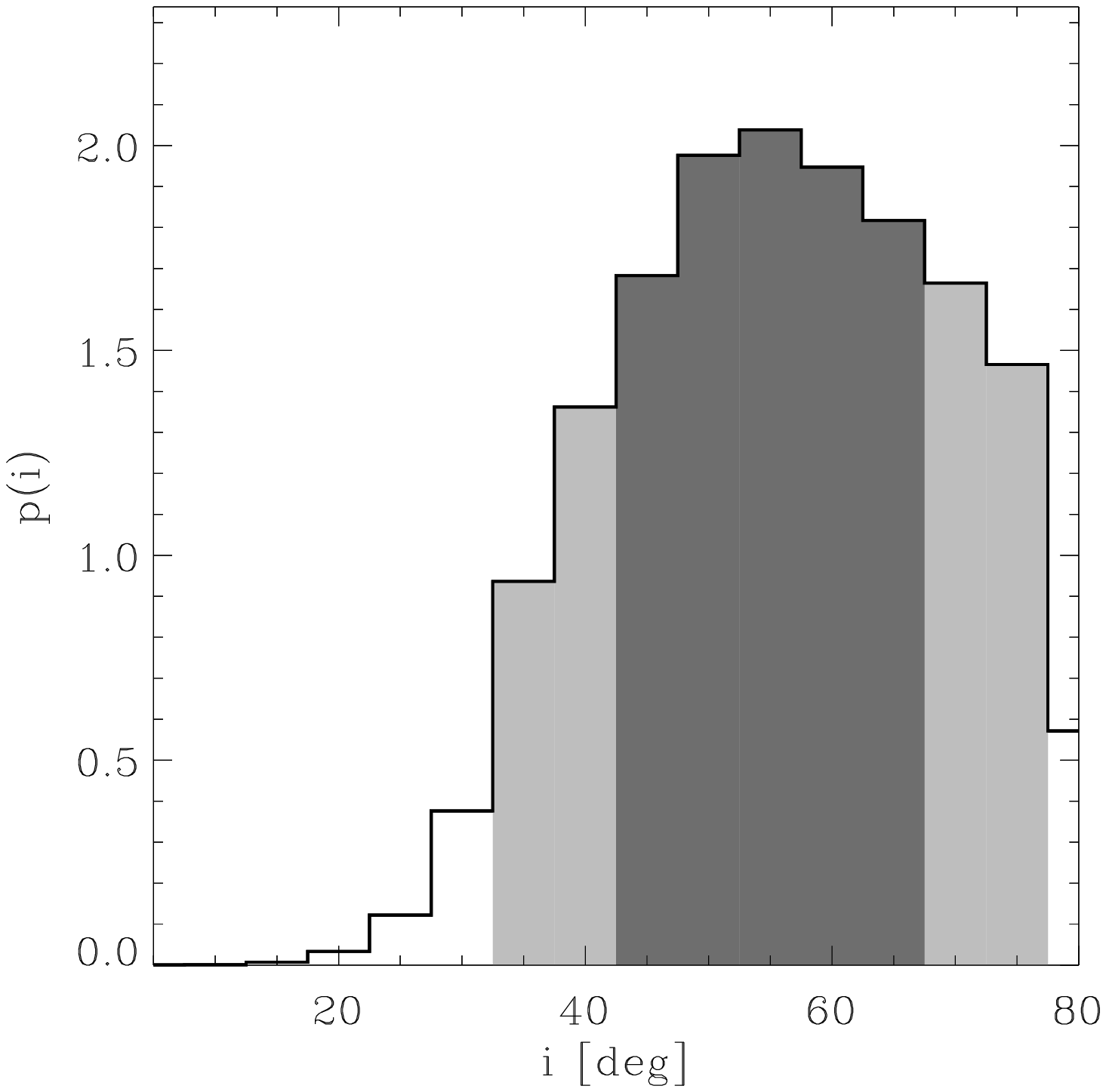}}
\centering{\includegraphics[width=0.33\textwidth]{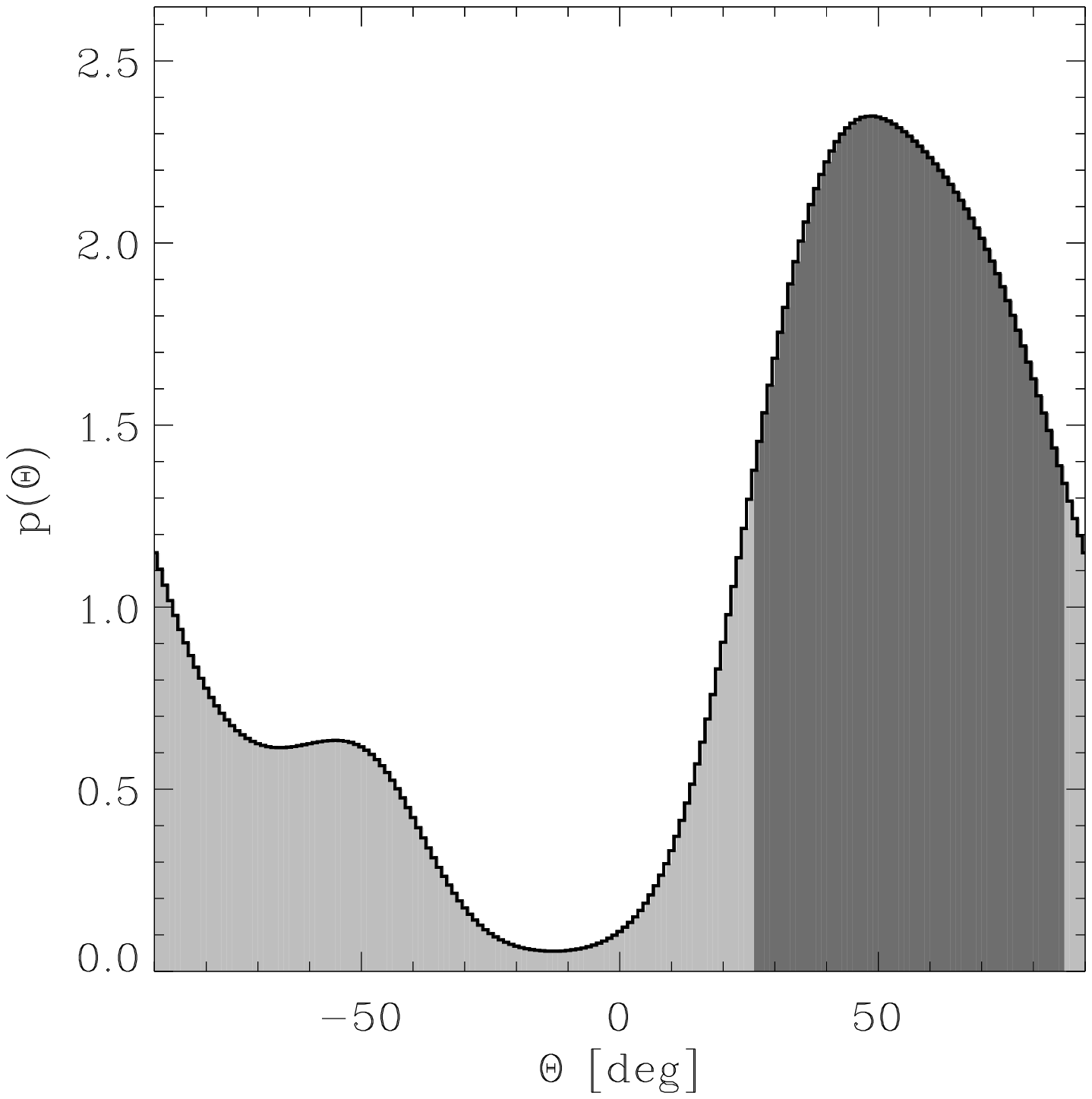}}
\caption[]{Same as Fig. \ref{modelA} assuming model B for $p(\xi)$.}
\label{modelB}
\end{figure*}

The NIR data we investigate here were taken in the $K_s$ band with the NIR camera CONICA and the
adaptive optics (AO) module NAOS (NACO) at the ESO VLT unit
telescope 4 (YEPUN) on Paranal, Chile\footnote{ Based on observations
at the Very Large Telescope (VLT) of the European Southern
Observatory (ESO) on Paranal in Chile; Programs:075.B-0093 and
271.B-5019(A).} on 13 June 2004 (start of observation: 07:20:02 UT).
For determining the linear polarization characteristics of a time-varying source, 
a Wollaston prism, installed on the NACO NIR camera, permits simultaneous measurements
of two orthogonal directions of the electric field vector. It is combined with a rotary
half-wave plate to allow for rapid alternating measurements of the electric field vector at different angles.  
Details of the data reduction and flux calibration is described in Eckart et al. (2006a), and Zamaninasab et al. (2010).

Figure~\ref{lag_obs1} shows light curves of a polarized flare from Sgr~A*
observed on 13 June 2004 in four polarization channels. 
There is a clear time-lag visible
between orthogonal channels which is bigger
than the sampling rate of the observations.
Furthermore, the light curve is continuous 
and there is no gap  according to the usual
sky background measurements at the time of the flare. 
Each channel shows an approximate Gaussian behaviour while clear time-lags between maxima 
(specially between orthogonal pairs) are visible.

We have calculated the time lag between orthogonal pair of channels for the original (Fig. \ref{lag_obs1} middle)
and first-order interpolated light curves (Fig. \ref{lag_obs1} bottom).
For the cross-correlation function we have followed Alexander (1997). 
Results of both methods are well in agreement with each other. 
This leads to the estimated values of
the time delays between orthogonal polarization channels 
 $\bar{\delta}_1=(4.20\pm1.00) \  \textrm{min}, \bar{\delta}_2=(2.12\pm1.00) \ \textrm{min}$ and a magnification
value of $\bar{\mu}=(2.98\pm1.00)$. The errors in these estimations derived from the standard 
deviation of $\chi^2$ fitting of Gaussian functions to the observed flux and cross-correlation functions.  
We must note that the sampling rate of this observation was $\sim 2 \ \textrm{min}$, so the 
level of confidence specially for the $\delta_2$ value is not high.
In order to improve these estimates 
we need high SNR images observed in polarimetry with higher time resolutions. 

Although the time resolution of the available data is not better than 1.5 minutes, it is good enough to rule
out some parts of the $a-i$ plane. Fig.~\ref{ai_sgr} shows two-dimensional marginalised probability densities $p(a,i)$
for the measured values of magnification and time delays. The Four different images correspond to different assumptions 
about the spatial distribution of the spots ($p(\xi)$). We see that while the most popular assumption
that spots are located around the ISCO (model A) results in $a=0.5^{+0.3}_{-0.4}$ and $i=50^{\circ+20^\circ}_{ -20^\circ}$, 
the more reasonable assumption which allows the spots to be created over a wider range of locations increases the 
estimated values for both inclination and black hole's spin. 

\begin{figure*}
\centering{\includegraphics[width=0.33\textwidth]{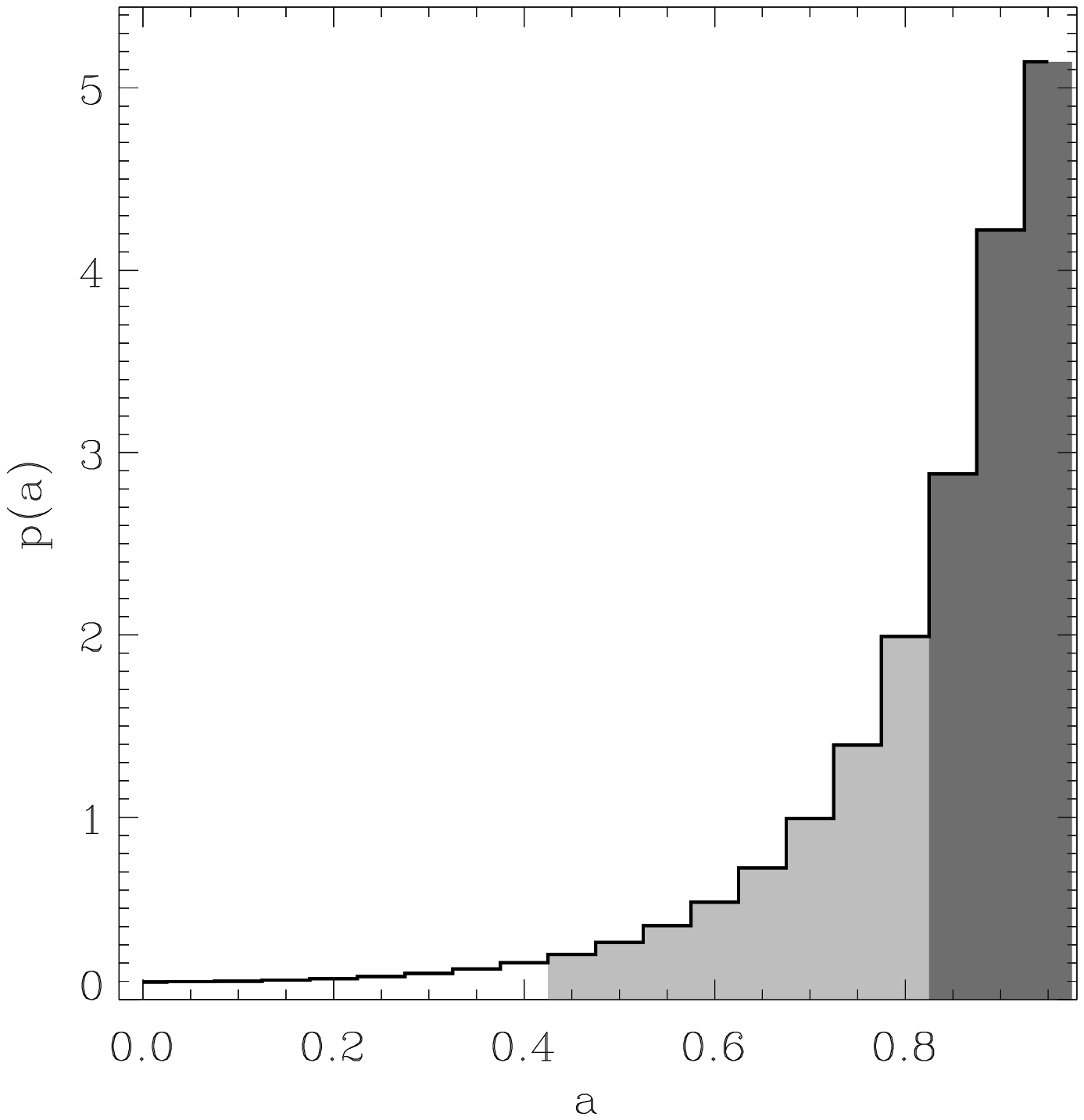}}
\centering{\includegraphics[width=0.33\textwidth]{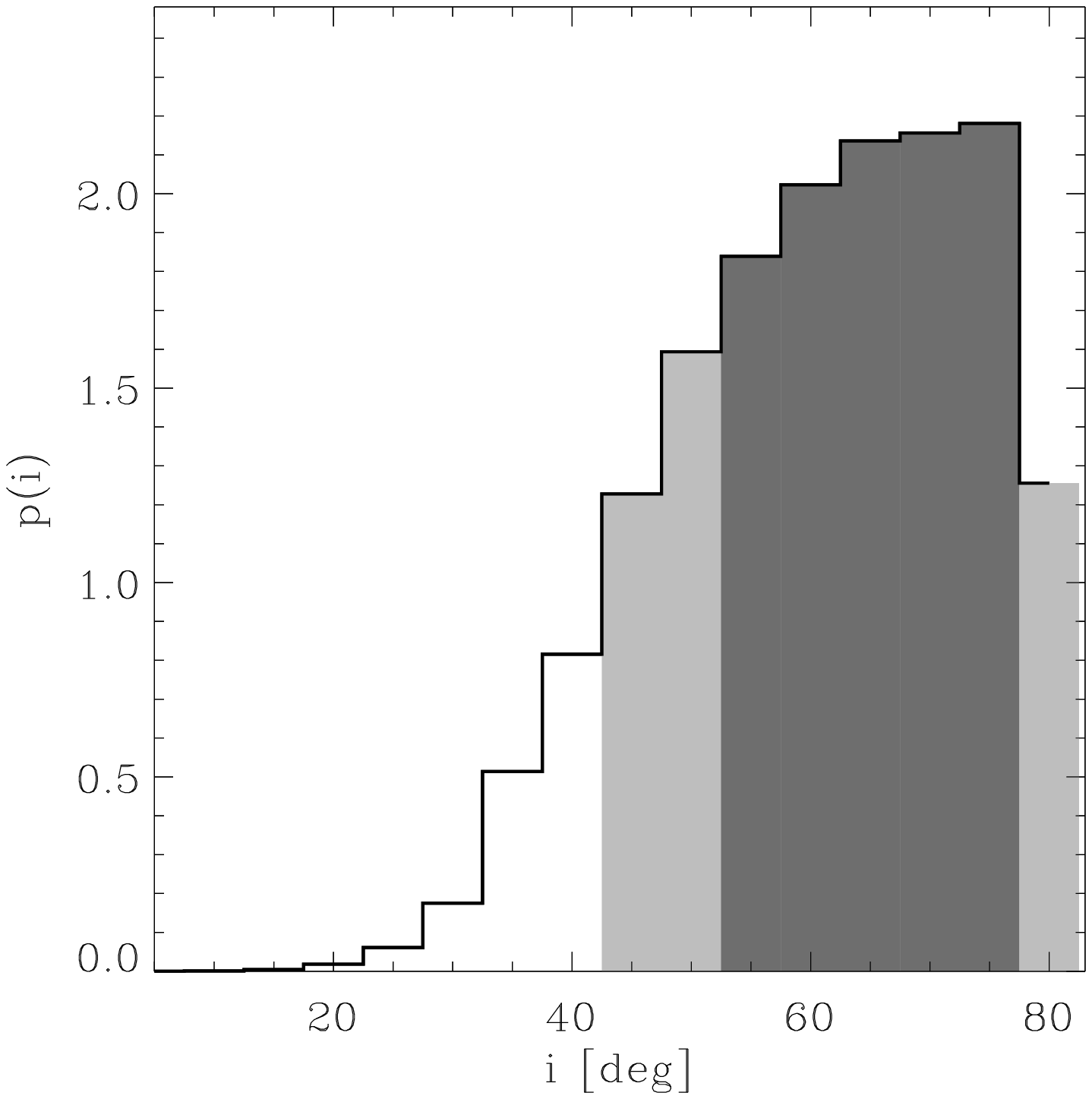}}
\centering{\includegraphics[width=0.33\textwidth]{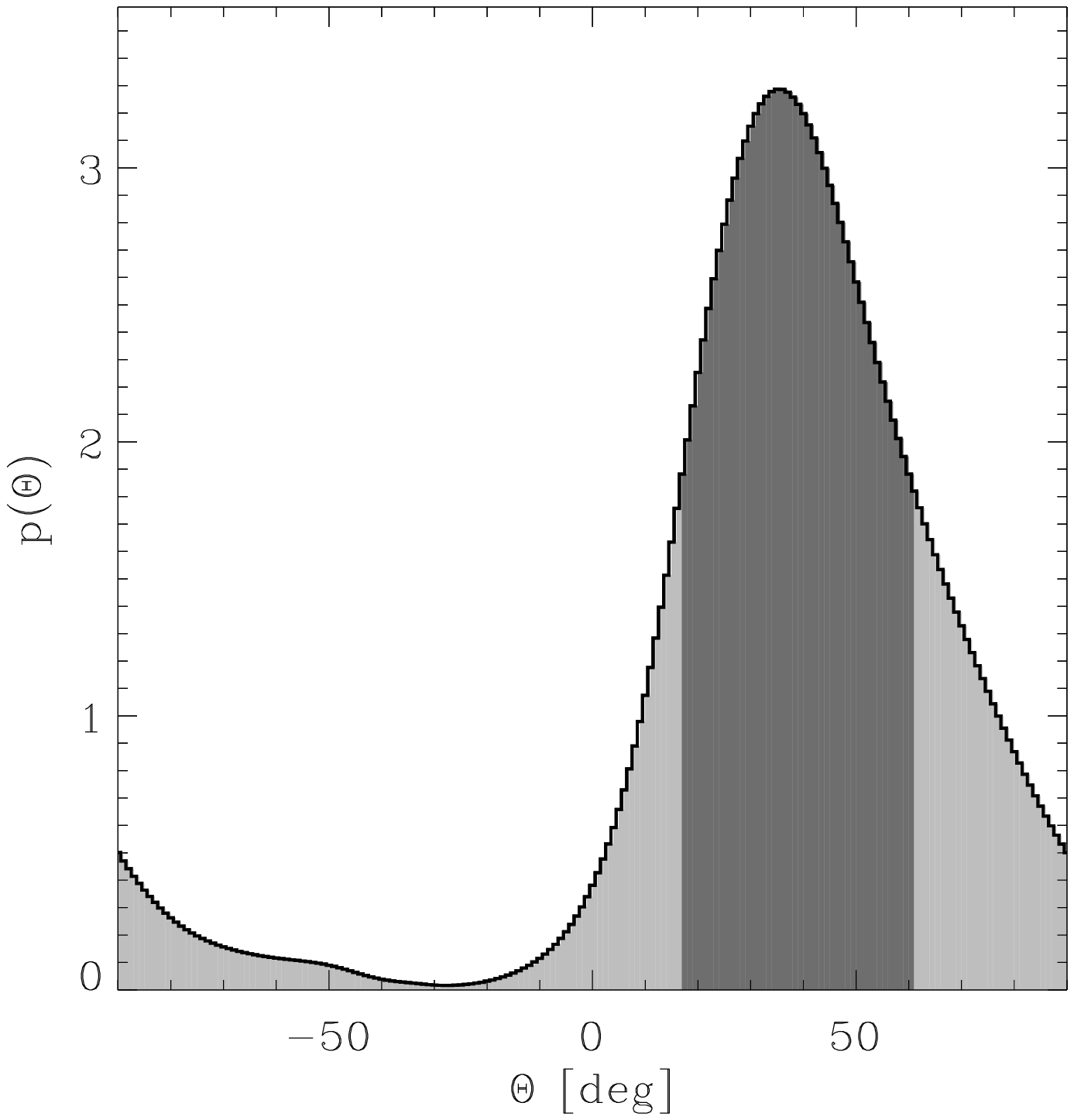}}
\caption[]{Same as Fig. \ref{modelA} assuming model C for $p(\xi)$.}
\label{modelC}
\end{figure*}
\begin{figure*}
\centering{\includegraphics[width=0.33\textwidth]{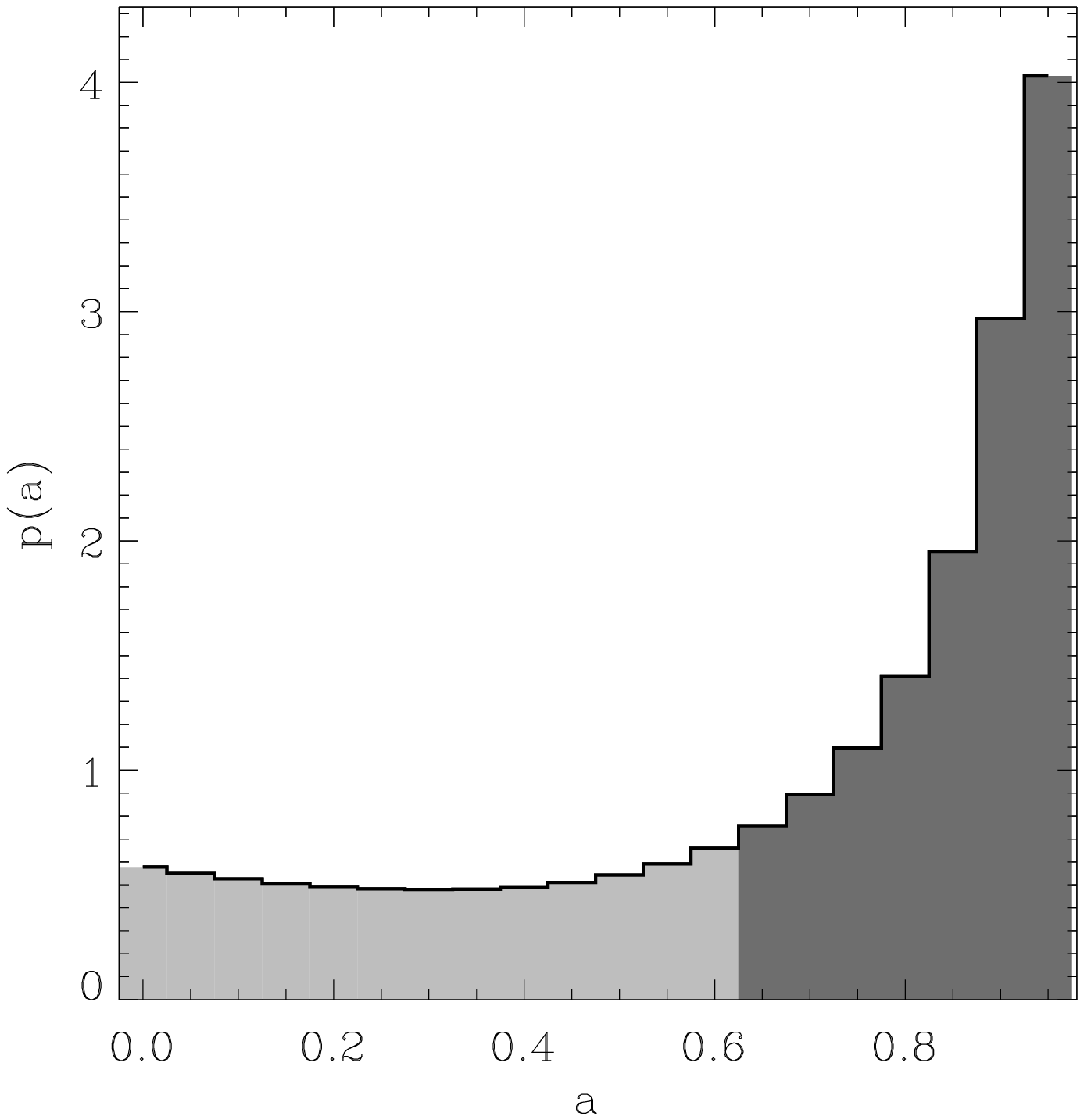}}
\centering{\includegraphics[width=0.33\textwidth]{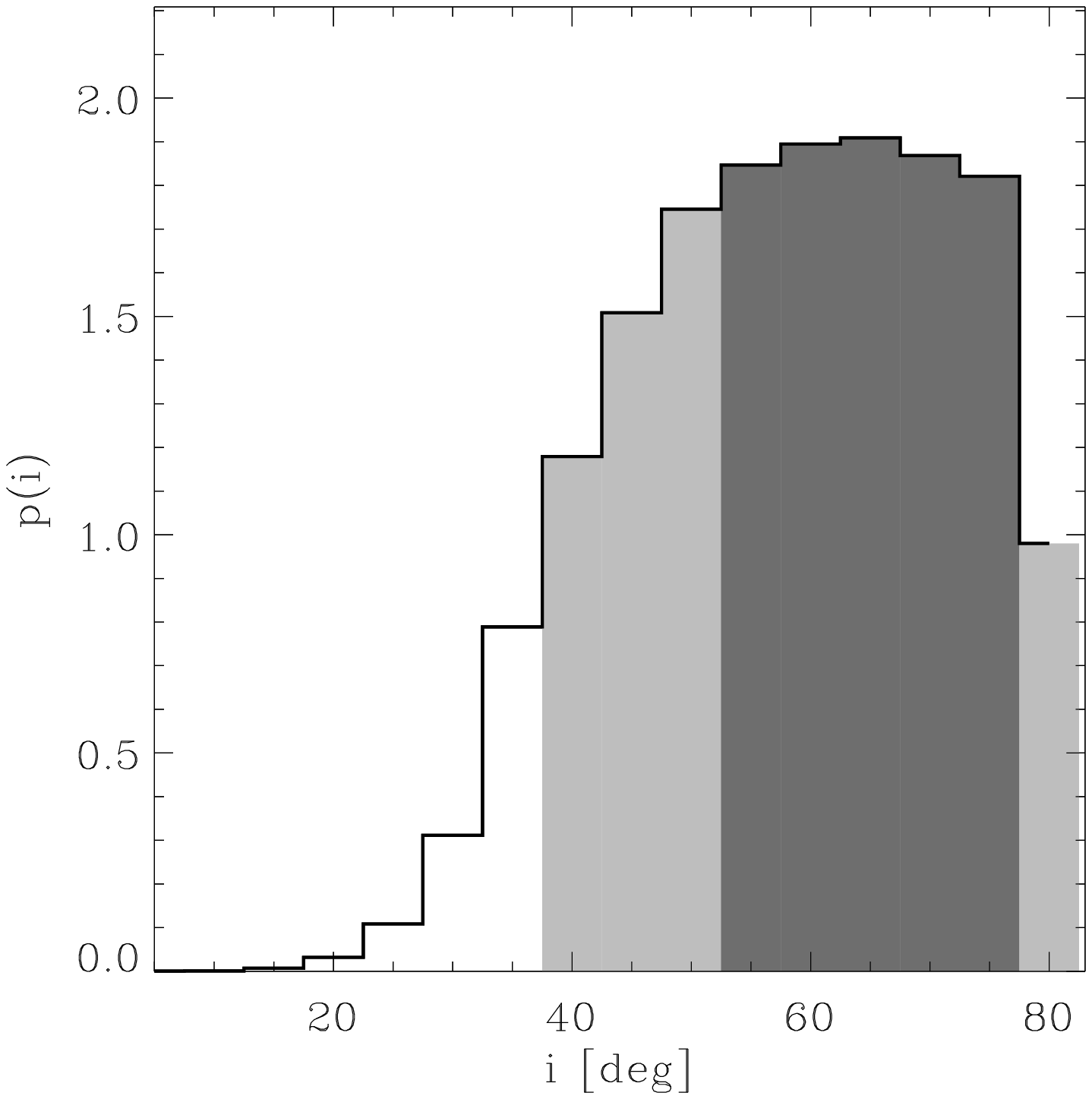}}
\centering{\includegraphics[width=0.33\textwidth]{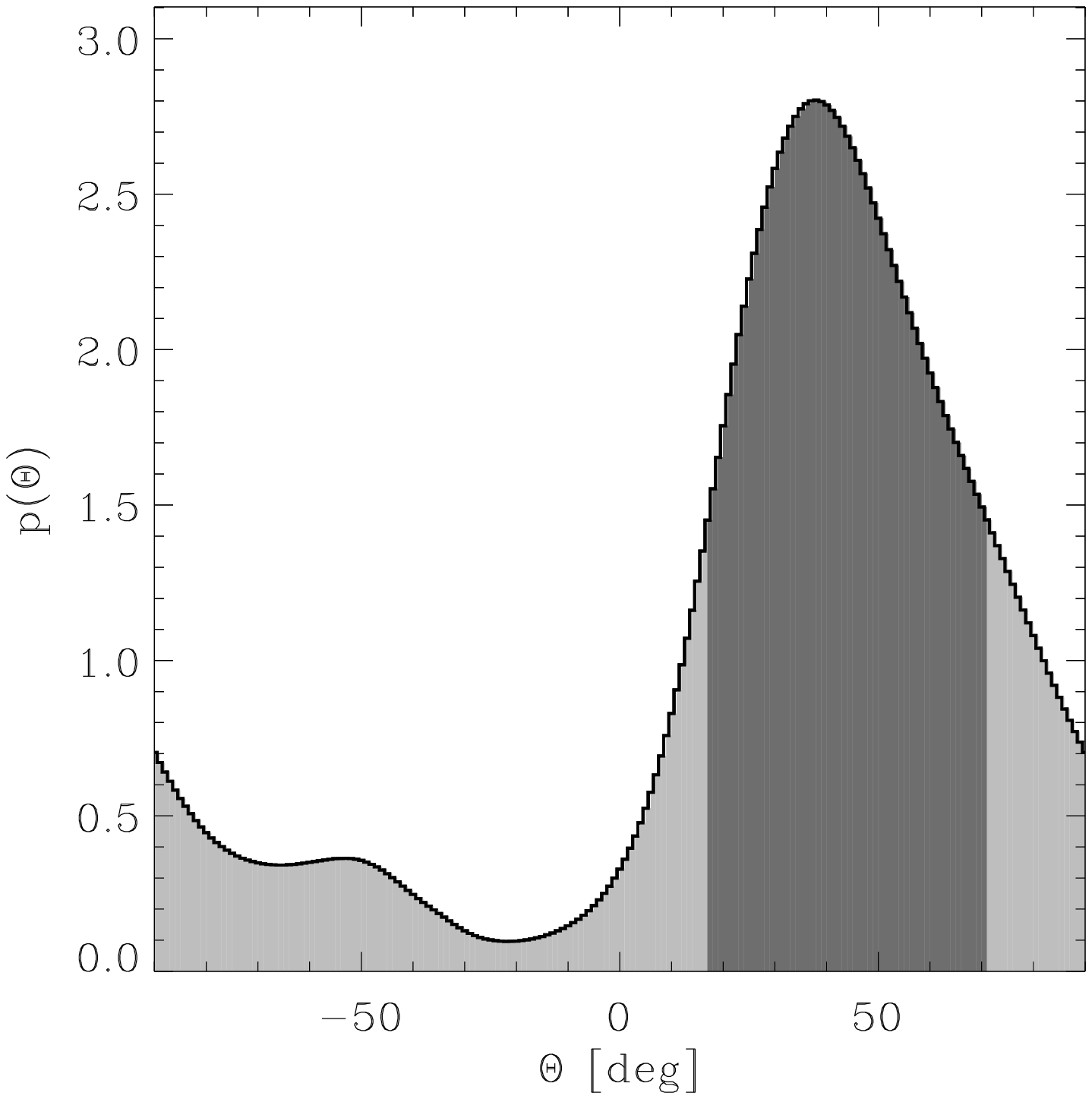}}
\caption[]{Same as Fig. \ref{modelA} assuming model D for $p(\xi)$.
}
\label{modelD}
\end{figure*}

Figs.~\ref{modelA}-\ref{modelD} show the fully marginalised probability distributions for different assumptions about $p(\xi)$, respectively.
  However, we must remind the reader that these parameter estimates are strongly correlated, and thus they should be used with caution. 
Here we again use the additive probability, $P(>p)$ to define the
$1\sigma$ and $2\sigma$ intervals, shown in the panels of Figs.~\ref{modelA}-\ref{modelD} as the dark and light shaded regions, respectively.

From the left panel of Fig.~\ref{modelA}, the most likely spin value is $a=0.45^{+0.2}_{0.15}$ ($1\sigma$ confidence). While we can rule out very low spins ($a\le0.1$)
at the $2\sigma$ level for our particular choice of $p(\xi)$, the spin is otherwise weakly constrained. Left panels of Figs.~\ref{modelA}-\ref{modelD} show 
that this estimation is highly dependant on the choice of $p(\xi)$ and has a tendency toward higher spin values for models B-D.

In contrast, the inclination is robustly limited towards moderate and high angles. The most likely inclination value 
is $i ={55^\circ}^{+20^\circ}_{-20^\circ}$ ($1\sigma$ confidence). It can be seen clearly in the central panels of 
Figs.~\ref{modelA}-\ref{modelD} that face-on geometries  ($i \le 30^\circ$) are convincingly ruled out. However, 
the general trend of all the considered models is toward highly inclined systems.

Interestingly, constraining the position angle shows that all the $p(\xi)$ assumptions 
result in a more or less similar values ($\theta=50^{\circ+30^\circ}_{-20^\circ}$) 
as can be seen in the right panels of Figs.~\ref{modelA}-\ref{modelD}. 

Our limits upon the black hole's spin are in good agreement with previous estimates by Huang et al. 2009, 
Dexter et al. 2010 and Moscibrodzka et al. 2009 while 
it contradicts the low spin values favoured by Broderick et al. 2009. However, we remind the 
reader again that the constraints on spin are not strong enough to rule out any possibility.

The limits upon the inclination are also in quite good agreement with previous efforts 
(Meyer et al. 2006, Markoff et al. 2007, Broderick et al. 2009, Falanga et al. 2009, Huang et al. 2009, 
Moscibrodzka et al. 2009 and Dexter et al. 2010). They all favour high inclinations using different approaches to the problem. 
Falanga et al. 2009, for example, find an inclination of about $77^\circ\pm10^\circ$ (similar to the results by Meyer et al. 2006) 
by fitting Sgr~A*'s flares with a model of hydrodynamic instabilities. On the other hand, Markoff et al. 2007 fit the long-wavelength 
observations with a different approach, a hydrodynamic jet. They also favour large inclinations ($i\ge 75^\circ$).

Unlike the inclination, our position angle estimate disagrees significantly with many previous efforts except 
Meyer et al. 2006 and Markoff et al. 2007. Our estimated position angle is marginally consistent with the 
second solution of Broderick et al. 2009 at the 1$\sigma$ level.

\begin{figure}
\centering{\includegraphics[width=0.48\textwidth]{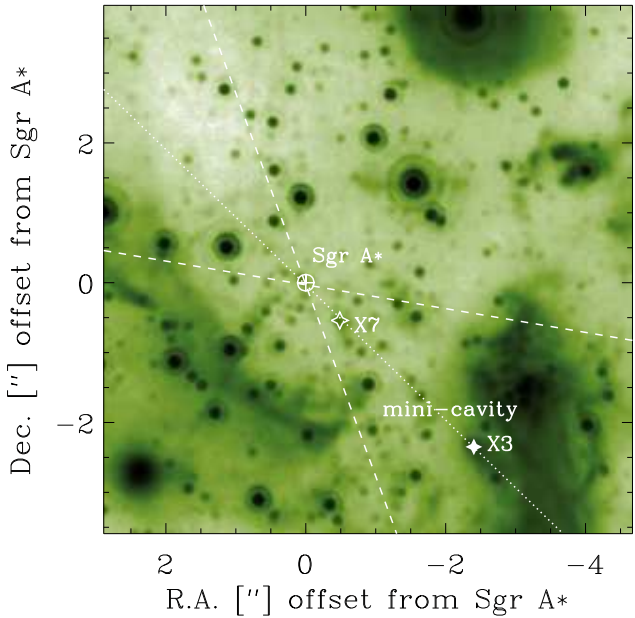}}
\caption[]{The estimated orientations of the Sgr~A*'s spin axis in
our sky from all the models of $p(\xi)$ ($\theta=50^{+30}_{-20}$) are in 
very well agreement with the study of Muzic et al. 2010 which relates the alignment of two bow shock sources
(X3 \& X7, marked with open and filled stars respectively) and mini-cavity as a possible hint for a wind from Sgr~A*.}
\label{gcwind}
\end{figure}

However, we have noticed that the our estimated $\theta$ coincides very well with the value derived by 
Mu\v{z}i\'{c} et al. 2010 where they used the structure of two bow shock stars (namely X3 and X7) 
in the vicinity of Sgr~A* to trace a possible outflow scenario (Fig.~\ref{gcwind}). 
Our best bet ($\theta=50^\circ$) exactly passes the positions of X3, X7 and the mini-cavity as shown in Fig.~\ref{gcwind}. 
It is proposed that the whole mini-cavity structure can be caused by wind from the direction of Sgr~A* implied by traces 
of hot gas and plasma shocks observed in this region 
(Yusef-Zadeh et al. 1990, Yusef-Zadeh \& Melia 1992, Yusef-Zadeh \& Wardle 1993 and Muzic et al. 2007).
At the same time, the results leave a lot of space for future improvements. 

\subsection{The case of RE~J1034+396}
\label{REJ}
The main advantage 
of this method is its applicability for a wide range of compact sources 
which show polarized variable emission in NIR and X-ray wavelengths. Since the current technical difficulties do not allow 
for polarimetric observations with time resolutions less than $\sim 30$ seconds (at least as far as it concerns VLT), 
the method is currently applicable to sources with masses greater than $\sim10^6 M_\odot$. This 
makes objects like RE~J1034+396  perfect candidates for testing  predictions of the hot spot model.  
 The recent
unambiguous discovery of one hour periodicity in the X-ray
emission light curve of this source is interpreted to be related to the 
ISCO frequency of this super-massive black hole (Gierli\'{n}ski et al. 2008).
Unfortunately, there is no polarimetric observation of this source available so far. The mass of the object
is also not well determined 
(the estimates vary from $6.3\times10^5$ to $3.6\times10^7$\solm).
This source looks to be a promising target for future X-ray polarimetry if its
mass estimation can be improved (using reverberation mapping for example).

\section{Summary and Conclusion}
\label{conclusion}
We have investigated the general relativistic effects on the polarization properties of 
the emission from an orbiting spot inside the accretion flow
around a black hole. We have described a method in which 
observing the time delay between the polarization channels in high-frequency
 regimes can lead to a constraint on the spin and inclination 
of the black hole. 
The method may also be applied to the X-ray domain for sources which are bright 
enough in polarized X-ray flux.
We applied the method to several hot spot models that
 sample a wide range of parameter space.
We have discussed how the probability for the spatial distribution
of the spots can affect our results. This shows the crucial need
for reliable 2 and 3-dimensional GRMHD simulations of accretion disks
concerning the exact location of the plunging region and 
the stress edge. 
The results of the application of the method to the available
 NIR polarimetric data of Sgr~A* is also 
presented. The results for this source are consistent with 
previous findings of an inclined accretion disk around a spinning black hole ($a\geq$0.4).
While with current VLT facilities the rate at which high S/N images in polarized flux density are taken can be increased significantly by a new method of observation, the next generations of NIR and X-ray polarimeters should be able to exploit this method as a powerful probe of the metric around black holes.

\section*{Acknowledgements}
Authors would like to thank the anonymous referee for his/her helpful comments on the paper. 
Part of this work was supported by the German \emph{Deut\-sche
For\-schungs\-ge\-mein\-schaft, DFG\/} via grant SFB 494. MZ, DK and MV-S 
are members of the International Max
Planck Research School (IMPRS) for Astronomy and Astrophysics at the
MPIfR and the Universities of Bonn and Cologne. NS
acknowledges support from the Bonn-Cologne Graduate School of Physics
and Astronomy (BCGS). RS acknowledges the Ram\'{o}n y 
Cajal program of the Spanish Ministry of Science and Innovation.
 VK and MD acknowledge the Czech Science
Foundation (ref. 205/07/0052).



\bsp

\label{lastpage}

\end{document}